\documentclass[mathleft
]{an}
\usepackage{graphicx}
\usepackage{times}
\begin{document}

\Pagespan{493}{}
\Yearpublication{2009}%
\Yearsubmission{2009}%
\Month{2}%
\Volume{330}%
\Issue{5}%
 \DOI{10.1002/asna.200911196}%

\title{Photometric monitoring of the young star Par 1724 in Orion\thanks{Based on 
observations obtained with telescopes of the University Observatory Jena, 
which is operated by the Astrophysical Institute of the Friedrich-Schiller-University;
a telescope of the University Observatory Munich on Mount Wendelstein,
the 0.9m ESO-Dutch telescope on La Silla, Chile, 
and with the All Sky Automated Survey (ASAS) project (www.astrouw.edu.pl/asas).}}

\author{R. Neuh\"auser\inst{1}\fnmsep\thanks{Corresponding author:
 {rne@astro.uni-jena.de}}
\and A. Koeltzsch\inst{1}
\and St. Raetz\inst{1}
\and T. O. B. Schmidt\inst{1}
\and M. Mugrauer\inst{1}
\and N. Young\inst{2}
\and F. Bertoldi\inst{3}
\and T. Roell\inst{1}
\and T. Eisenbeiss\inst{1}
\and M. M. Hohle\inst{1,4}
\and M. Va{\v n}ko\inst{1}
\and C. Ginski\inst{1}
\and W. Rammo\inst{1}
\and M. Moualla\inst{1}
\and\\ C. Broeg\inst{5}
}
\titlerunning{Photometric monitoring of Par 1724}

\institute{Astrophysikalisches Institut und Universit{\"a}ts-Sternwarte Jena, 
Schillerg{\"a}{\ss}chen 2-3, D-07745 Jena, Germany
\and
Jodrell Bank Centre for Astrophysics, University of Manchester, Alan-Turin Building, Manchester, M13 9PL, UK
\and
Argelander-Institut f\"ur Astronomie, Universit\"at Bonn, Auf dem H\"ugel 71, D-53121 Bonn, Germany 
\and
MPI Extraterrestrische Physik, Giessenbachstrasse, D-85740 Garching, Germany
\and
Space Research and Planetary Sciences, Physikalisches Institut, 
University of Bern, Sidlerstra{\ss}e 5, 3012 Bern, Switzerland
}

\received{2009 Feb 20}
\accepted{2009 Mar 17}
\publonline{2009 May 30}


\abstract{We report new photometric observations of 
the ${\sim\! 200\,000}$ year old naked weak-line run-away T Tauri star 
Par 1724, located north of the Trapezium cluster in Orion.
We observed in the broad band filters $B$, $V$, $R$, and $I$ using the 90\,cm Dutch telescope
on La Silla, the 80\,cm Wendelstein telescope, and a 25\,cm telescope of the 
University Observatory Jena in Gro\ss schwabhausen near Jena.
The photometric data in $V$ and $R$ are consistent with a $\sim\! 5.7$ day rotation
period due to spots, as observed before between 1960ies and 2000.
Also, for the first time, we present evidence for a long-term 9 or 17.5 year
cycle in photometric data ($V$ band) of such a young star, 
a cycle similar to that to of the Sun and other active stars.
}

\maketitle

\section{Introduction: Par 1724}

About 400 years ago, Galileo could observe the rotation of a star
due to spots and wrote: {\it I am at last convinced that the spots are objects close
to the surface of the solar globe ... also that they are carried around the 
Sun by its rotation} (Galileo 1613). 
Later on, Schwabe (1843) published in this journal the ${2\!\times\! 11}$ year
Sun spot cycle. In the last few decades, the rotational periods of many stars have 
been determined from periodic photometric variability due to surface spots.
Also, several old stars (${\ge\! 4.6}$ Gyr) were shown to display cycles with similar length,
observed often as variability in the Ca H \& K emission
(e.g. Baliunas et al. 1995), sometimes also in broad photometric bands (e.g. Alekseev 2005);
only rarely, stars younger than 1 Gyr were shown to display
cycles (Baliunas et al. 1995; Alekseev 2005).

The K0 pre-main sequence star Par 1724 (or P1724) in Orion is listed as star number 1724
in Parenago (1954), it is located at $\alpha = 5^{\rm h} 35^{\rm m} 4.21^{\rm s}$
and $\delta = -5^{\circ} 8^{\prime} 13.2^{\prime \prime}$ (J2000.0),
i.e. just $15^{\prime}$ north of the Trapezium cluster;
this star is also called V1321 Ori (Kazarovets \& Samus 1997),
JW 238 (Jones \& Walker 1988), or HBC 452 (Herbig \& Bell 1988).

Par 1724 is one of the most active and variable young stars known,
see e.g. Neuh\"auser et al. (1998) for an extensive observational study:
It shows a photometric rotation period of $\sim\! 5.7$ days (seen so far in all
data sets from 1968 to 1997); this period is also evident in the radial
velocity data; the photometric variability is mostly due to a large
polar spot detected indirectly by Doppler imaging; 
the star also shows strong and variable H$\alpha$ and
X-ray emission as well as other youth indicators (Li 6708\,\AA~absorption); 
according to its luminosity (${\sim\! 49}$~L$_{\odot}$) and temperature
(5250 K for K0 spectral type) and comparison with evolutionary models, 
it is ${\sim\! 200\,000}$ years old with a mass of ${\sim\! 3}$~M$_{\odot}$;
there is no evidence for multiplicity nor for infrared
excess emission or circumstellar material;
while its mean radial velocity is fully consistent
with membership to the Orion Trapezium cluster, its proper motion
is somewhat different, but consistent with having been ejected from the
Trapezium towards the north some 100\,000 years ago;
hence, Par 1724 is a run-away weak-line naked T Tauri star; 
see Neuh\"auser et al. (1998) for details.
The rotational period of ${\sim\! 5.7}$ days has also been found by
Cutispoto et al. (1996).

The previous photometry resulted in the following average magnitudes
(with ranges given in parentheses), data are from Neuh\"auser et al. (1998)
and literature given therein: \\[3pt]
$B = 11.87 \pm 0.25$ mag (from 11.41 to 12.33 mag), \\
$V = 10.72 \pm 0.13$ mag (from 10.44 to 10.99 mag), \\
$R = 9.81 \pm 0.17$ mag (from 9.44 to 10.18 mag), and \\
$I = 9.08 \pm 0.16$ mag (from 8.71 to 9.44 mag). \\[3pt]
The star may also show a long-term change of its optical magnitude, 
namely getting fainter by $\sim\! 0.2$ mag in $V$ over 40 years, 
see Fig. 5 in Neuh\"auser et al. (1998), but with long data gaps.

In this work, we observed Par 1724 again in the 
Bessel $B$, $V$, $R$, and $I$ filters as well as in the Gunn $i$ filter:
We took new images in March 1998 with the 90\,cm Dutch Telescope on La Silla;
then more new images in late 2004 and early 2005 with the 90\,cm telescope
on Mount Wendelstein observatory of Ludwig-Maximilians University
of Munich (USM), located in the southern German Alps;
and finally many new images in 2007 and 2008 with 
the 25\,cm (11 inch) Cassegrain-Teleskop-Kamera (CTK) 
installed piggy back on the tube of the 90\,cm telescope at the
University Observatory Jena near Gro\ss schwabhausen (GSH),
see Mugrauer (2009) for details on CTK.

We present the observations and data reduction in Sect. 2,
and the results of absolute and relative photometry in Sects. 3
and 4, respectively. 
In Sect. 5, we also study again the long-term photometric
behaviour of Par 1724 using all data available so far
as listed in Neuh\"auser et al. (1998), Cutispoto (1998),
Cutispoto et al. (1998, 2001, 2003), this work, plus
also $V$-band data from the All Sky Automated Survey (ASAS) 
project available online.
We summarize the results in Sect. 6.

\section{Observation and data reduction}

For the observations at the Dutch Telescope, 
we used the Tektronix TK512CB grade 1 
thinned back-illuminated CCD chip, $512 \times 512$ pixels,
each 27 $\mu$m in size, to give a field of 
view of $3.77^{\prime} \times 3.77^{\prime}$;
see the ESO 90\,cm Dutch Telescope Users Manual 
for more information (www.eso.org).

For the observations at the USM Wendelstein 80\,cm telescope, 
we used the CCD camera MONICA, equipped with a 1k Tektronix chip 
and Johnson-Bessel $B$, $V$, $R$, and $I$ filters, mounted at the Cassegrain focus.

For the GSH/CTK observations, we use an optical CCD 
camera IMG 1024S from Finger Lake Instrumentation
with a $2.2065 \pm 0.0008^{\prime \prime}$ per pixel 
as pixel scale and
a $37.7^{\prime} \times 37.7^{\prime}$ field of view,
$1024 \time 1024$ pixels with 24 $\mu$m each;
see Mugrauer (2009) for details on telescope and CCD.

We present the observations log in Table 1.

\begin{table}
\caption{Observations log for Par 1724.}
\begin{tabular}{lcrrc} \hline\noalign{\smallskip}
\quad Date     & Filter & No. of & Exposure & Abs./Rel. \\ 
\quad of Night &        & Images & [s]~~~~   & Photom. \\[1.5pt] \hline\noalign{\smallskip}
\multicolumn{5}{c}{Dutch telescope observations} \\ \hline
28 Feb 1998       & $V$   & 8            & 15 to 40 & abs. \& rel. \\
to                & $R$   & 8            & 10 to 20 & abs. \& rel. \\
1 Mar 1998        & $i$   & 11           & 15 to 30 & abs. \& rel. \\ \hline\noalign{\smallskip}
\multicolumn{5}{c}{Wendelstein telescope observations} \\ \hline
24/25 Nov 2004    & $B$      & 6            & 60 to 120 & abs. \& rel. \\
               & $V$      & 5            & 15 to 25 & abs. \& rel. \\
               & $R$      & 4            & 5 to 20  & abs. \& rel. \\
               & $I$      & 15           & 1 to 20  & abs. \& rel. \\
25/26 Nov 2004    & $B$      & 7            & 40 to 75 & abs. \& rel. \\
               & $V$      & 10           & 1 to 60  & abs. \& rel. \\
               & $R$      & 5            & 10 to 15 & abs. \& rel. \\
               & $I$      & 10           & 1 to 60  & abs. \& rel. \\
7/8 Dec 2004     & $B$      & 5            & 15 to 50 & abs. \& rel. \\
               & $V$      & 5            & 10 to 25 & abs. \& rel. \\
               & $R$      & 4            & 20       & abs. \& rel. \\
               & $I$      & 5            & 10 to 20 & abs. \& rel. \\
10/11 Dec 2004    & $B$      & 4            & 75 to 200 & abs. \& rel. \\
               & $V$      & 6            & 12 to 25  & abs. \& rel. \\
               & $R$      & 3            & 7 to 12   & abs. \& rel. \\
               & $I$      & 10           & 12 to 30  & abs. \& rel. \\
29/30 Jan 2005    & $B$      & 5            & 100       & abs. \& rel. \\ 
               & $V$      & 3            & 40 to 100 & abs. \& rel. \\
               & $R$      & 8            & 1 to 20   & abs. \& rel. \\
               & $I$      & 6            & 10 to 20  & abs. \& rel. \\ \hline\noalign{\smallskip}
\multicolumn{5}{c}{University Observatory Jena (GSH)} \\ \hline 
15/16 Mar 2007 & $V$      & 1             & 60  & \\ 
               & $R$      & 3             & 60  & relative \\
               & $I$      & 3             & 60  & relative \\ \hline
27/28 Mar 2007 & $V$      & 3             & 60  & relative \\ 
               & $R$      & 3             & 60  & relative \\ \hline
4/5 Apr 2007   & $B$      & 3             & 120 & relative \\
               & $V$      & 3             & 60  & relative \\
               & $R$      & 3             & 60  & relative \\
               & $I$      & 3             & 60  & relative \\ \hline
5/6 Apr 2007   & $B$      & 12            & 120 & relative \\
               & $V$      & 12            & 60  & relative \\
               & $R$      & 9             & 60  & relative \\
               & $I$      & 9             & 60  & relative \\ \hline
11/12 Apr 2007 & $B$      & 4             & 120 & relative \\
               & $V$      & 3             & 60  & relative \\
               & $R$      & 4             & 60  & relative \\
               & $I$      & 4             & 60  & relative \\ \hline
12/13 Apr 2007 & $B$      & 3             & 120 & relative \\
               & $V$      & 3             & 60  & relative \\
               & $R$      & 3             & 60  & relative \\
               & $I$      & 3             & 60  & relative \\ \hline
13/14 Apr 2007 & $B$      & 4             & 120 & relative \\
               & $R$      & 6             & 60  & relative \\
               & $I$      & 6             & 60  & relative \\ \hline
14/15 Apr 2007 & $B$      & 4             & 120 & rel. \& abs. \\
               & $V$      & 8             & 60  & rel. \& abs. \\
               & $R$      & 8             & 60  & rel. \& abs. \\
               & $I$      & 8             & 60  & rel. \& abs. \\ \hline
23/24 Oct 2008 & $V$      & 5             & 60  & rel. \& abs. \\
               & $R$      & 5             & 60  & rel. \& abs. \\
               & $I$      & 5             & 50  & rel. \& abs. \\ \hline
\end{tabular}
\end{table}

For all science frames obtained, we also took several darks with the
same exposure time; and for all filters used, we also obtained
several sky flat fields (and again several darks with the same exposure time).
We then took the median of the darks for each exposure time
and subtracted the corresponding medium dark from the flat field frame
for each filter. The bias level is included in the dark frames.
Then, we normalized the flat fields and took the mean for each filter.
Finally, we could subtract the darks from the science frames and
divide them by the respective normalized mean flat field. All 
this was done with IRAF and MIDAS. 

We show a typical image of Par 1724 in Fig. 1 (three color
image with Orion nebula) and Fig. 2 ($R$-band image of central 
part with comparison stars indicated).

\begin{figure*}
\centering
\includegraphics[width=16cm]{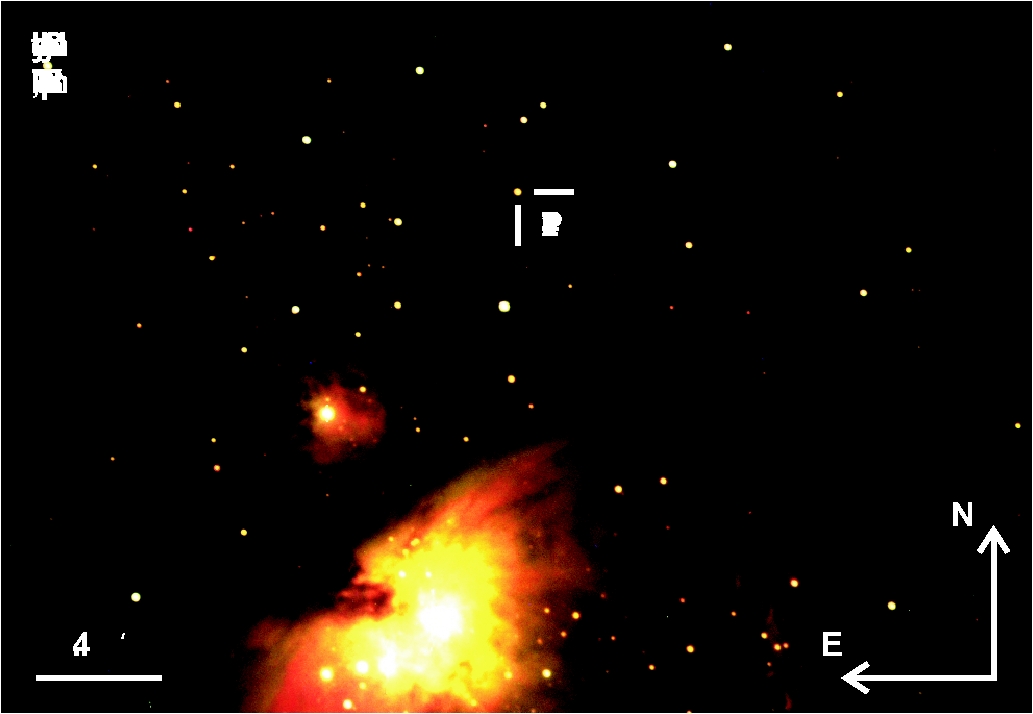}
\caption{Color composite image from bands $BV\!R$ of our GSH/CTK 
images of the star Par 1724 (marked in the upper central part)
and the Orion nebula at the bottom. (Obtain good (e)ps files for this and 
other figures from www.astro.uni-jena.de/Observations/gsh/gsh$\_$papers.htm)}
\end{figure*}

\begin{figure}
\centering
\includegraphics[width=8cm]{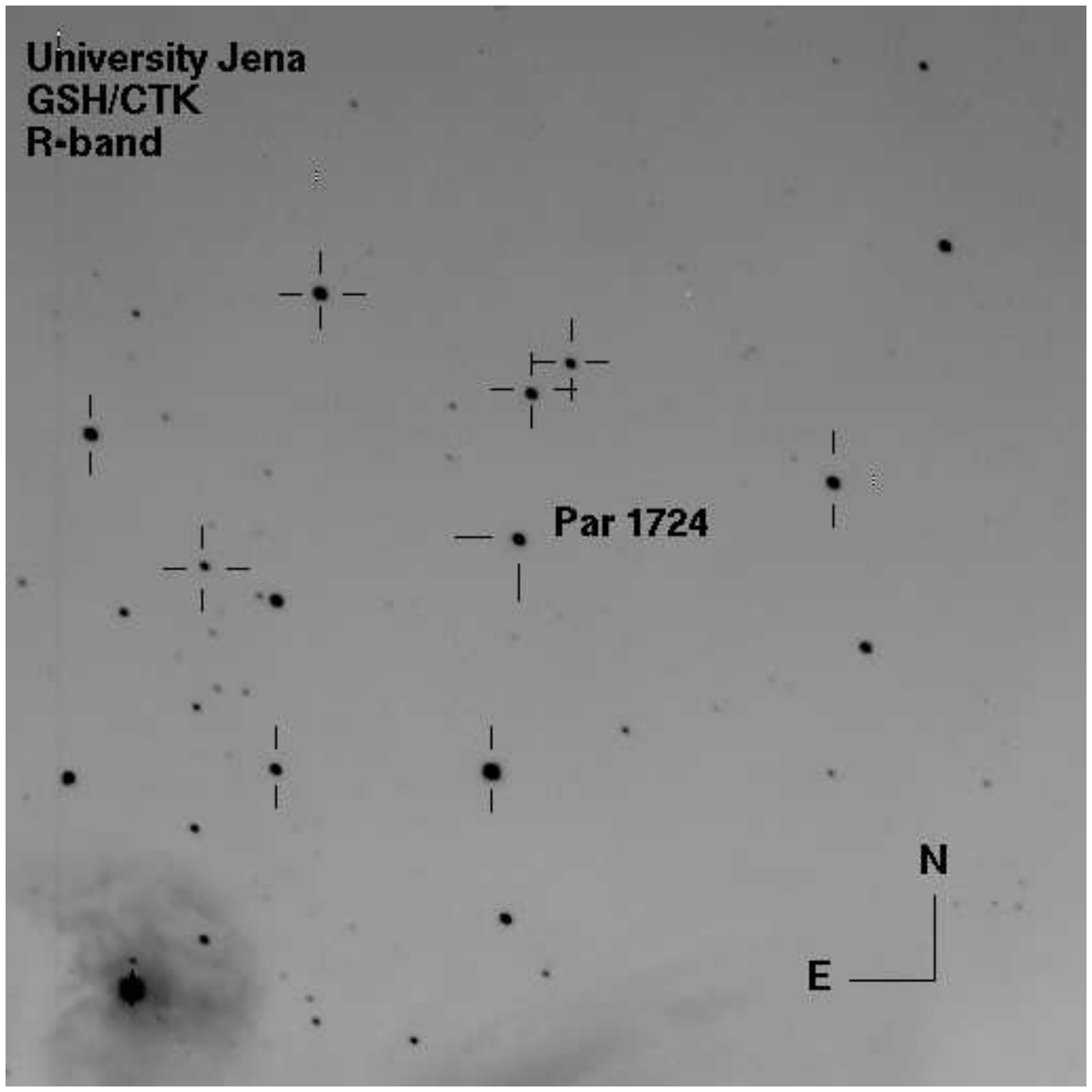}
\caption{Central part of one of our GSH/CTK $R$-band images 
of the star Par 1724 (60\,s exposure) in the center 
with the comparison stars used for relative photometry 
also marked; those stars indicated with four tick marks each
are the four comparison stars used for absolute photometry
for the Wendelstein data (Table 2);
field size shown is $18.3^{\prime}\! \times\! 18.3^{\prime}$.}
\end{figure}

\section{Absolute photometry}

During the night 28 Feb to 1 March 1998 on La Silla, we also observed
several photometric standard stars from Landolt (1992).

Also, in the nights 14/15 April 2007 and 23/24 Oct 2008 at GSH, 
we observed photometric standard stars from Landolt (1992) and
Galadi-Enriquez et al. (2000) at two or more different airmasses, 
see Figs. 3 to 5 for those fields and Table 2 for the known
magnitudes of the standard stars.

\begin{figure}
\centering
\includegraphics[width=8cm]{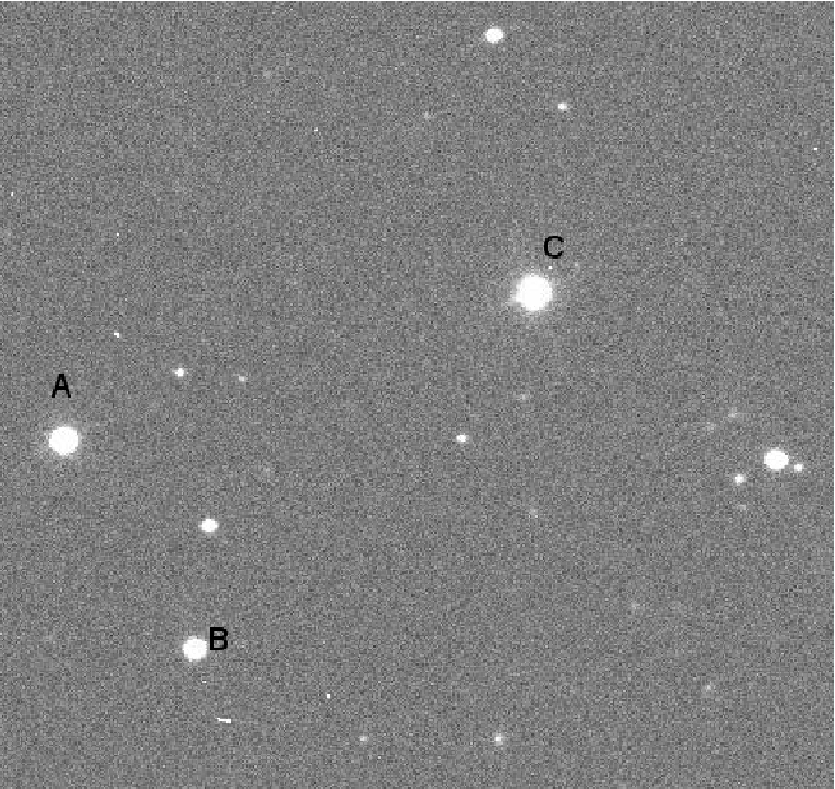}
\caption{Part of one of our GSH/CTK $R$-band images of the Landolt
standard star field PG 1047+003 (40\,s exposure) with the three
standard stars A, B, and C marked;
field size shown is $3.8^{\prime}\! \times\! 3.8^{\prime}$;
North up, East left.}
\end{figure}

\begin{figure}
\centering
\includegraphics[width=8cm]{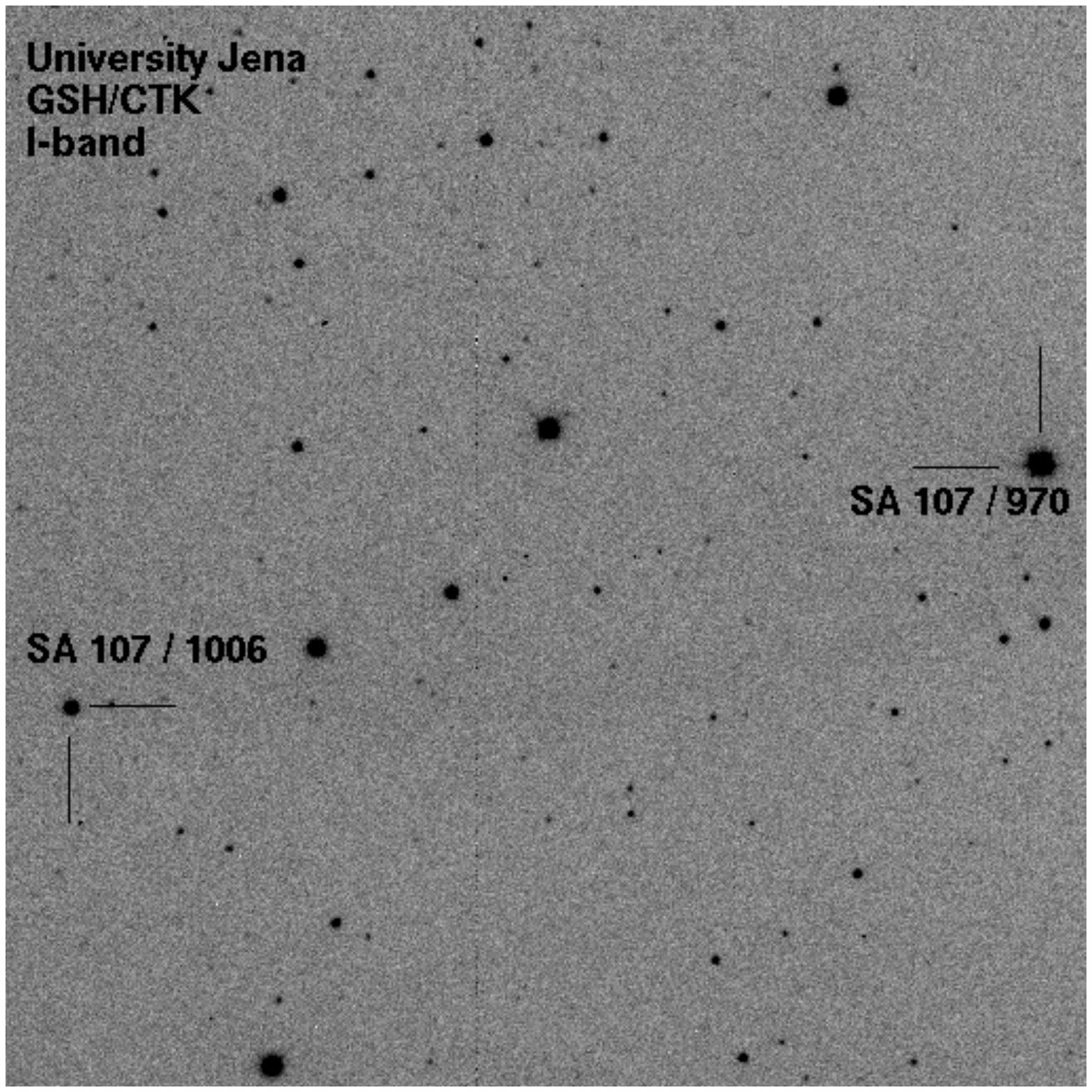}
\caption{Part of one of our 0.9\,m-Dutch telescope $I$-band images 
of the Landolt standard star field SA 107 (15\,s exposure) 
with the two standard stars 970 and 1006 marked; 
field size shown is $18.3^{\prime}\! \times\! 18.3^{\prime}$;
North up, East left.}
\end{figure}

\begin{figure}
\centering
\includegraphics[width=8cm]{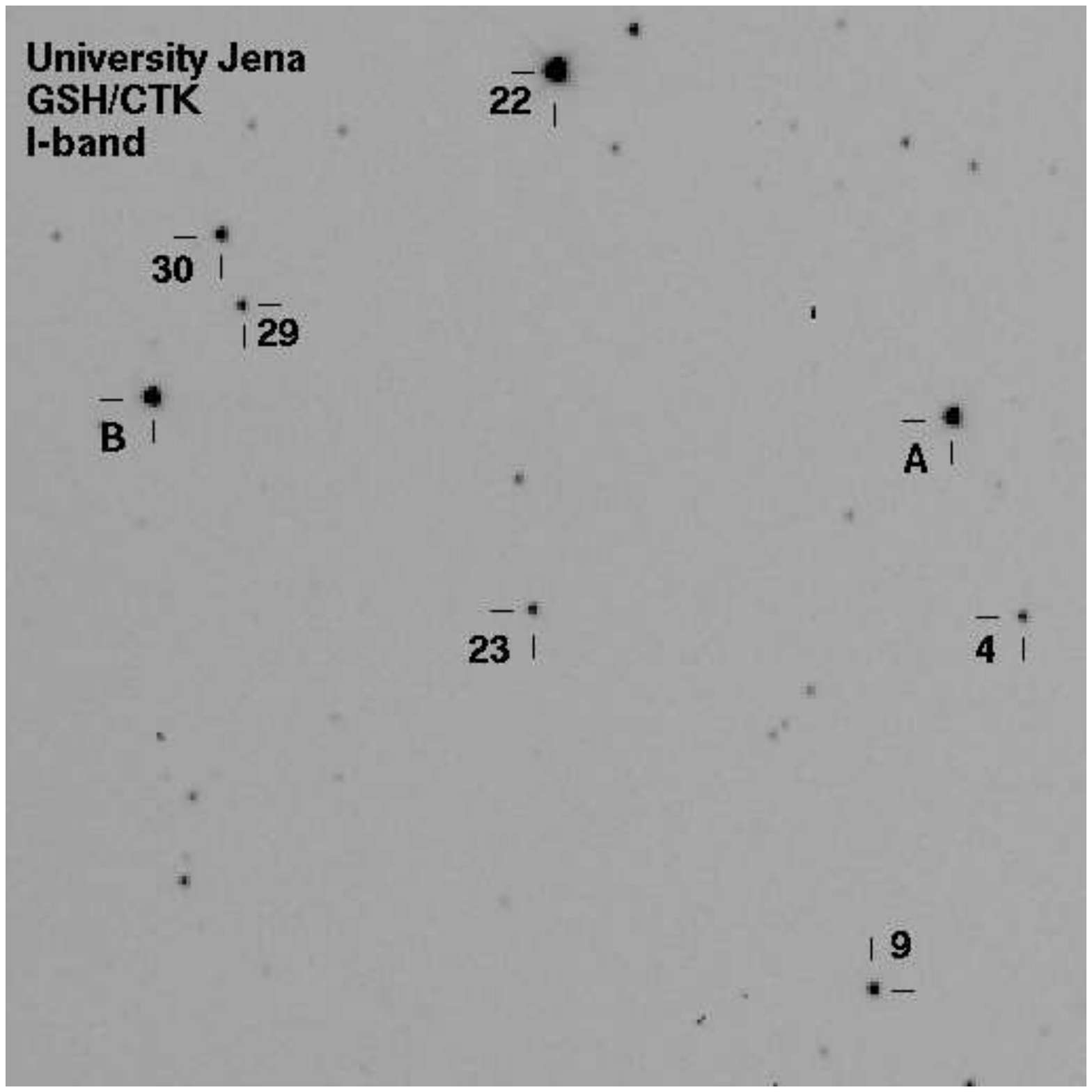}
\caption{Part of one of our GSH/CTK $I$-band images of the 
Galadi-Enriques standard star field number 1 around
Landolt standard stars SA 4428 (labelled A) and
SA 44 113 (labelled B) 
with stars used for relative photometry as marked
by letters and numbers (60\,s exposure); 
field size shown is $9.2^{\prime}\! \times\! 9.2^{\prime}$;
North up, East left.}
\end{figure}

\begin{table*}
\caption{Photometric standard stars observed.}
\begin{tabular}{lllllll} \hline\noalign{\smallskip}
Date of Night & Standard Stars &\qquad\quad $B$ &\qquad\quad $V$ &\qquad\quad $R$ &\qquad\quad $I$ & Ref. (*) \\[1.5pt] \hline\noalign{\smallskip}
28 Feb to   & PG 1047+003 A & & $13.512 \pm 0.0046$ & $13.0900 \pm 0.0053$ & $12.6720 \pm 0.0058$ & L 1992 \\
1 March     & PG 1047+003 B & & $14.751 \pm 0.0050$ & $14.3600 \pm 0.0055$ & $13.9870 \pm 0.0099$ & L 1992 \\
1998        & PG 1047+003 C & & $12.453 \pm 0.0093$ & $12.075 \pm 0.01$ & $11.716 \pm 0.010$ & L 1992 \\[1.5pt] \hline\noalign{\smallskip}
24 Nov 2004 & BD$-05^{\circ}$1310 & $10.997 \pm 0.025$ & $10.458 \pm 0.031$ & $10.029 \pm 0.045$ & $\enspace9.859 \pm 0.403$ & VizieR \\
to          & BD$-05^{\circ}$1309 & $11.340 \pm 0.020$ & $10.991 \pm 0.009$ & $10.680 \pm 0.049$ & $10.439 \pm 0.105$ & VizieR \\
29 Jan 2005 & BD$-05^{\circ}$1316 &\enspace $9.383 \pm 0.025$ &\enspace $9.348 \pm 0.038$ & $\enspace9.425 \pm 0.063$ & $\enspace9.418 \pm 0.051$ & VizieR \\
            & Par 2020 & $12.843 \pm 0.127$ & $11.930 \pm 0.066$ & $11.127 \pm 0.133$ & $10.864 \pm 0.162$ & VizieR  \\[1.5pt] \hline\noalign{\smallskip}
14/15 Apr   & SA 107 970  & $12.535 \pm 0.0076$ & $10.939 \pm 0.0074$ & $\enspace9.797 \pm 0.0077$ & $\enspace8.365 \pm 0.0084$ & L 1992 \\
2007        & SA 107 1006 & $12.478 \pm 0.0014$ & $11.712 \pm 0.0010$ & $11.270 \pm 0.0013$ & $10.849 \pm 0.0014$ & L 1992 \\[1.5pt] \hline\noalign{\smallskip}
23/24 Oct  & A  & & $11.329 \pm 0.002$ & $10.935 \pm 0.0022$ & $10.565 \pm 0.0028$ & G-E 2000 \\
2008       & B  & & $11.713 \pm 0.006$ & $11.046 \pm 0.0067$ & $10.484 \pm 0.0078$ & G-E 2000 \\
           & 5  & & $13.844 \pm 0.003$ & $13.489 \pm 0.0042$ & $13.155 \pm 0.0058$ & G-E 2000 \\
           & 9  & & $13.099 \pm 0.007$ & $12.744 \pm 0.0092$ & $12.387 \pm 0.0081$ & G-E 2000 \\
           & 22 & & $10.187 \pm 0.003$ & $\enspace9.6040 \pm 0.0095$ & $\enspace9.085  \pm 0.0058$ & G-E 2000 \\
           & 23 & & $13.569 \pm 0.005$ & $13.268 \pm 0.0058$ & $12.978 \pm 0.0078$ & G-E 2000 \\
           & 29 & & $13.563 \pm 0.008$ & $13.215 \pm 0.010 $ & $12.897 \pm 0.0094$ & G-E 2000 \\ 
           & 30 & & $12.747 \pm 0.008$ & $12.368 \pm 0.0094$ & $12.747 \pm 0.0089$ & G-E 2000 \\[1.5pt] \hline
\end{tabular}
\\[3pt]
Remarks: $B$ and $V$ in the Johnson system, $R$ and $I$ in the Cousin system;
(*) L 1992 is Landolt (1992), and G-E 2000 is Galadi-Enriques et al. (2000);
VizieR: median from data collected from VizieR (mostly from Morel \& Magnenat 1978).
\end{table*}

During the observing nights at Wendelstein, we did not observe 
official photometric standard stars in addition to Par 1724, 
because we were originally only interested in relative photometry
of Par 1724 and other stars.
However, several nights were good enough for absolute photometry, so that we can
try to use constant bright stars in the Par 1724 fields for absolute
photometric calibration. We use the most constant and brightest four comparison stars
determined as constant comparison stars for the relative photometry in
the next section (see Fig. 2). 
The mean of the $BV\!RI$ magnitudes found with VizieR
(mostly from Morel \& Magnenat 1978) are listed in Table 2.

We obtained aperture photometry with IRAF 
(for Dutch and Wendelstein telescope data)
and MIDAS (for GSH data) for all images obtained.
In the Par 1724 field, we obtained photometry for Par 1724 itself
and several comparison stars around Par 1724.
For the standard stars, we used the same aperture size as for Par 1724,
but different sizes in different nights depending on the FWHM.

For nights with absolute photometric conditions (clear and cloudless),
we then used photometric standard stars to determine
the atmospheric extinction and zero points for each filter
using the well-known equation
\begin{equation}
m = c + m_{\rm instr} - k\! \cdot\! Y
\end{equation}
for apparent magnitude $m$ (known for standard stars
or to be obtained for science targets),
instrumental magnitude,
detector zero point $c$, atmospheric extinction $k$,
and airmass $Y$. 
Using the aperture photometry of the standard stars, we can then
obtain the apparent magnitudes for $B$, $V$, $R$, and $I$ for Par 1724,
see Table 3, for all three telescopes used. The Gunn $i$ filter data 
can be transformed to standard $I$ using Jordi et al. (2006).
We will use the new absolute photometric data in Sect. 5 below
to investigate possible long-term brightness changes.

\begin{table*}
\caption{New absolute photometry for Par 1724$^1$.}
\begin{tabular}{lclll} \hline\noalign{\smallskip}
\qquad  Date           & UT          &\quad Zero Point$^2$  &\quad Extinction & \qquad Result$^3$  \\
\qquad MJD            & [hh:mm]     &\quad\ \ c [mag]    &\quad\ \ k [mag]    &\qquad [mag]      \\[1.5pt] \hline\noalign{\smallskip}
1 Mar 1998     & 00:44--03:15 & $22.155 \pm 0.028$ & $0.158 \pm 0.008$ & $V = 10.833 \pm 0.014$ \\
and            & 00:51--03:12 & $22.073 \pm 0.016$ & $0.073 \pm 0.004$ & $R = 10.076 \pm 0.008$ \\
MJD = 50873.1  & 00:56--03:22 & $21.485 \pm 0.033$ & $0.057 \pm 0.005$ & $I = 9.265 \pm 0.015$ \\ \hline\noalign{\smallskip}
24/25 Nov 2004 & 22:09--03:06 & $20.77 \pm 0.13$ & $0.265 \pm 0.064$ & $B = 11.872 \pm 0.06$ \\
and            & 22:14--03:10 & $22.01 \pm 0.12$ & $0.221 \pm 0.059$ & $V = 10.453 \pm 0.044$ \\
MJD = 53334.5  & 21:54--03:04 & $21.82 \pm 0.30$ & $0.09  \pm 0.15$  & $R = 9.656 \pm 0.013$ \\
               & 22:19--03:14 & $20.53 \pm 0.35$ & $0.09  \pm 0.18$  & $I = 9.14 \pm 0.058$ \\ \hline\noalign{\smallskip}
25/26 Nov 2004 & 21:37--03:09 & $20.85 \pm 0.11$ & $0.28  \pm 0.053$ & $B = 11.822 \pm 0.026$ \\
and            & 21:35--03:04 & $21.99 \pm 0.11$ & $0.179 \pm 0.050$ & $V = 389 \pm 0.013$ \\
MJD = 53335.0  & 21:32--03:01 & $21.93 \pm 0.21$ & $0.127 \pm 0.098$ & $R = 9.638 \pm 0.021$ \\
               & 21:30--02:13 & $20.49 \pm 0.31$ & $0.04  \pm 0.15$  & $I = 9.100 \pm 0.017$ \\ \hline\noalign{\smallskip}
7/8 Dec 2004   & 20:36--03:05 & $19.817 \pm 0.093$ & $0.058 \pm 0.035$ & $B = 11.87 \pm 0.21$ \\ 
and            & 20:42--03:07 & $20.627 \pm 0.089$ & $0.121 \pm 0.032$ & $V = 10.43 \pm 0.21$ \\ 
MJD = 53347.0  & 20:44--03:09 & $19.98 \pm 0.15$ & $0.357 \pm 0.062$ & $R = 9.67 \pm 0.41$ \\
               & 20:45--03:11 & $18.66 \pm 0.23$ & $0.367 \pm 0.095$ & $I = 9.15 \pm 0.41$ \\ \hline\noalign{\smallskip}
10/11 Dec 2004 & 21:58--02:36 & $21. 05 \pm 0.11$ & $0.422 \pm 0.045$ & $B = 12.16 \pm 0.20$ \\ 
and            & 21:52--02:33 & $22.077 \pm 0.095$ & $0.285 \pm 0.044$ & $V = 10.667 \pm 0.062$ \\
MJD = 53350.0  & 21:50--02:28 & $21.98 \pm 0.22$ & $0.17  \pm 0.10$  & $R = 9.73 \pm 0.029$ \\
               & 21:16--02:24 & $20.62 \pm 0.34$ & $0.11  \pm 0.17$  & $I = 9.243 \pm 0.011$ \\ \hline\noalign{\smallskip}
29 Jan 2005    & 18:00--23:31 & $22.21 \pm 0.21$ & $1.092 \pm 0.093$ & $B = 12.07 \pm 0.14$ \\
and            & 18:03--23:37 & $23.43 \pm 0.17$ & $1.021 \pm 0.070$ & $V = 10.609 \pm 0.092$ \\
MJD = 53399.9  & 17:32--23:40 & $23.38 \pm 0.16$ & $0.952 \pm 0.067$ & $R = 9.74 \pm 0.14$ \\
               & 18:07--23:42 & $21.99 \pm 0.22$ & $0.879 \pm 0.095$ & $I = 9.23 \pm 0.13$ \\ \hline\noalign{\smallskip}
14 Apr 2007    & 19:05--19:14 & $18.379 \pm 0.073$  & $0.35 \pm 0.15$     & $B = 11.3 \pm 0.3$ \\
and            & 19:15--19:39 & $19.356 \pm 0.051$  & $0.2556 \pm 0.0034$ & $V = 10.462 \pm 0.020$ \\
MJD = 54204.6  & 19:21--19:45 & $19.324 \pm 0.020$  & $0.211 \pm 0.017$   & $R = 9.809 \pm 0.033$ \\
               & 19:27--19:51 & $18.619 \pm 0.069$  & $0.074 \pm 0.034$   & $I = 9.235 \pm 0.077$ \\ \hline\noalign{\smallskip}
24 Oct 2008    & 01:07--01:37 & $18.939 \pm 0.067$  & $0.267 \pm 0.080$   & $V = 10.50 \pm 0.11$ \\
and            & 01:08--01:39 & $18.906 \pm 0.055$  & $0.158 \pm 0.073$   & $R = 9.745 \pm 0.091$ \\
MJD = 54763.1  & 01:13--01:40 & $18.215 \pm 0.12$   & $0.050 \pm 0.073$   & $I = 9.06 \pm 0.13$ \\ \hline
\end{tabular}
\\[3pt]
$^1$: from absolute photometric nights only; $^2$: given for that band and normalized to 1\,s exposure;
$^3$: in most cases, error-weighted mean values are given for several exposures, so that the
final magnitude error can be slightly smaller than the error in the zero point.
\end{table*}

\begin{table}[!h]
\caption{All new photometric data for Par 1724$^*$.}
\begin{tabular}{lllll} \hline\noalign{\smallskip}
\quad MJD  & $B$ [mag] & $V$ [mag] & $R$ [mag] & $I$ [mag] \\[1.5pt] \hline\noalign{\smallskip} 
\multicolumn{5}{c}{La Silla Dutch 90 cm telescope:} \\ \hline\noalign{\smallskip}
50873.031   &        &  10.856 &        &  9.280 \\
50873.032   &        &  10.851 &        &  9.277 \\
50873.033   &        &  10.847 &        &  9.279 \\
50873.036   &        &         & 10.078 & \\
50873.037   &        &         & 10.081 & \\
50873.038   &        &         & 10.080 & \\
50873.039   &        &         &        &  9.280 \\
50873.040   &        &         &        &  9.277 \\
50873.041   &        &         &        &  9.279 \\
50873.041   &        &         &        &  9.277 \\
50873.068   &        &  10.840 &        & \\
50873.069   &        &  10.836 &        & \\
50873.071   &        &         & 10.074 &  9.260 \\
50873.072   &        &         & 10.074 &  9.257 \\
50873.073   &        &         &        &  9.262 \\
50873.131   &        &         & 10.069 & \\
50873.132   &        &         & 10.075 & \\
50873.133   &        &         & 10.077 & \\
50873.134   &        &  10.812 &        & \\
50873.135   &        &  10.813 &        & \\
50873.136   &        &  10.811 &        & \\
50873.137   &        &         &        &  9.254 \\
50873.138   &        &         &        &  9.247 \\
50873.139   &        &         &        &  9.257 \\
50873.140   &        &         &        &  9.258 \\ \hline\noalign{\smallskip} 
\multicolumn{5}{c}{Wendelstein 80 cm telescope} \\ \hline\noalign{\smallskip}
53334.5     &  11.87 &  10.453 & 9.656  &  9.14 \\
53335.0     &  11.82 &  10.389 & 9.638  &  9.100 \\
53347.0     &  11.87 &  10.43  & 9.67   &  9.15 \\
53350.0     &  12.16 &  10.667 & 9.73   &  9.243 \\
53399.9     &  12.07 &  10.609 & 9.74   &  9.23 \\ \hline\noalign{\smallskip} 
\multicolumn{5}{c}{University Observatory Jena (GSH)} \\ \hline\noalign{\smallskip}
54175.0     &        &  10.330 & 9.778  &  9.274 \\
54186.7     &        &  10.356 & 9.821  & \\
54194.7     &  11.46 &  10.481 & 9.816  &  9.151 \\
54195.7     &  11.43 &  10.455 & 9.907  &  8.985 \\
54201.7     &  11.54 &  10.399 & 9.942  &  9.116 \\
54202.7     &  11.45 &  10.369 & 9.893  &  9.177 \\
54203.7     &  11.4  &         & 9.841  &  9.157 \\
54204.7     &  11.3  &  10.462 & 9.809  &  9.235 \\
54763.1     &        &  10.50  & 9.745  &  9.06 \\ \hline
\end{tabular}
\\[3pt]
$^*$ Data from both absolute photometric nights (as in Table 3) and
also other nights close to the absolute photometric nights, for
which we can deduce the photometry from the values in the
absolute photometric nights and stable standard stars.
\end{table}

\section{Relative photometry: rotation period}

For the data obtained in 2007, all within one 
month (nine nights from 15 March to 14 April),
we tried to find a periodicity in the photometry. We first searched for constant,
similarly bright, nearby comparison stars for Par 1724; 
after the first iteration, we omitted variable stars;
the remaining bright, constant comparison stars are shown in Fig. 2.
The brightness variations of those comparison stars have
standard deviations of $\pm 0.01$ mag or less;
two of them are listed as {\em variable stars} in Simbad,
but their variability amplitude was very small or
negligible during our observations;
the light curve of one of them is shown in Fig. 6.
The variability amplitude of Par 1724 is relatively large
with several tenth of a mag (Neuh\"auser et al. 1998, and this work).

\begin{figure}
\includegraphics[width=5cm,angle=270]{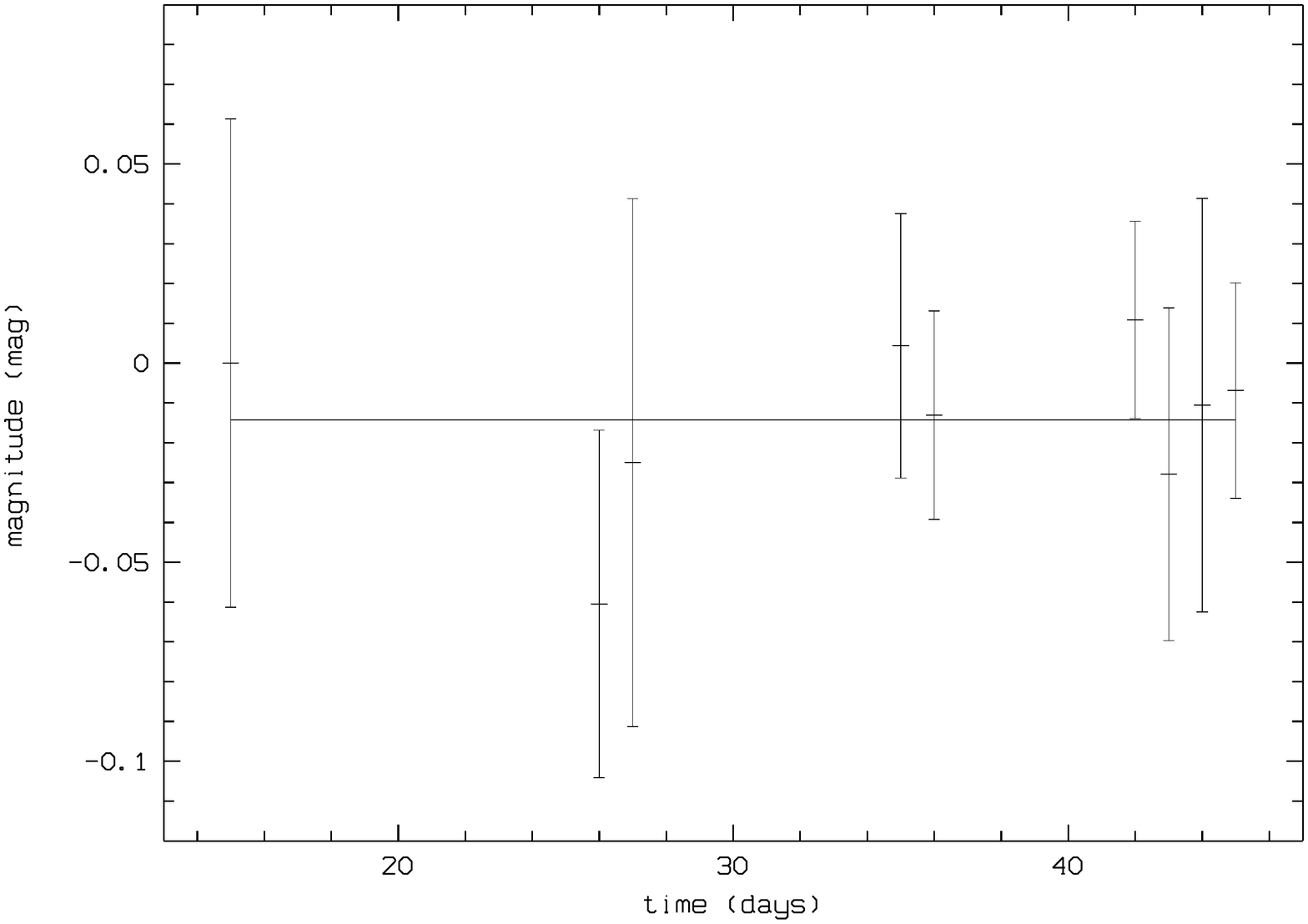}
\caption{We see no significant variations in the comparison stars,
one of which is plotted here (BD\,$-5^{\circ}$1316, relative 
magnitude versus observing day).
Standard deviations of our comparison stars are $\pm 0.01$ mag or less,
which is roughly the amplitude of the error bars achieved
in the Par 1724 data shown below.
These data were obtained with GSH/CTK in March and April 2007.
Relative magnitudes are plotted versus observing time
(in days since 1 February 2007).}
\end{figure}

We determined the photometric magnitude of Par 1724 relative to those eight
constant comparison stars, done for the bands $BV\!RI$. 
Then, we calculated the mean of the relative magnitude changes
between Par 1724 and each comparison star between the first night
and any other night.
We then searched for periodicity signals in the data using the standard
methods string length (Lafler \& Kinman 1965; Burke et al. 1983;
Dworetsky 1983; Broeg et al. 2005), Lomb-Scargle (Scargle 1982;
Horne \& Baliunas 1986; Broeg et al. 2007), 
and a Fourier analysis (Lenz \& Breger 2005).
Signals with low false-alarm probability are found 
only in the $V$ and $R$ bands (Figs. 7 and 8 for the $V$ band,
Figs. 9 and 10 for the $R$ band);
the data in the $B$ and $I$ bands are to noisy,
amplitudes are too small, or not enough data are available.

In both the $R$  and $V$ band, the string length, Lomb-Scargle,
and Fourier analysis give best periods of ${\sim\! 5.6}$ to 5.7 days.
The false-alarm probabilities are below 0.001.
Phase-folded light curves in $V$ and $R$ pass a visual inspection test best
for 5.7 days as period 
(and slightly less well for both 5.6 and 5.8 day periods). 
Taking all these tests into consideration, 
we obtain a rotational period of 5.7 days.
The final phase-folded light curves for the $V$ and $R$ bands 
with 5.7 day period are displayed in Figs. 11 and 12, respectively.

We can confirm a period of $\sim\! 5.7$ days already found by
Cutispoto et al. (1996) and Neuh\"auser et al. (1998) in earlier data,
the apparent period has not changed, a spot is still 
or again present, we see no indication for differential rotation.
The variability amplitude (peak-to-peak in sine wave fitted) obtained in 
our data in 2007 is $\Delta R \simeq 0.15$ mag and $\Delta V \simeq 0.2$ mag.

\begin{figure}
\includegraphics[width=7cm,angle=0]{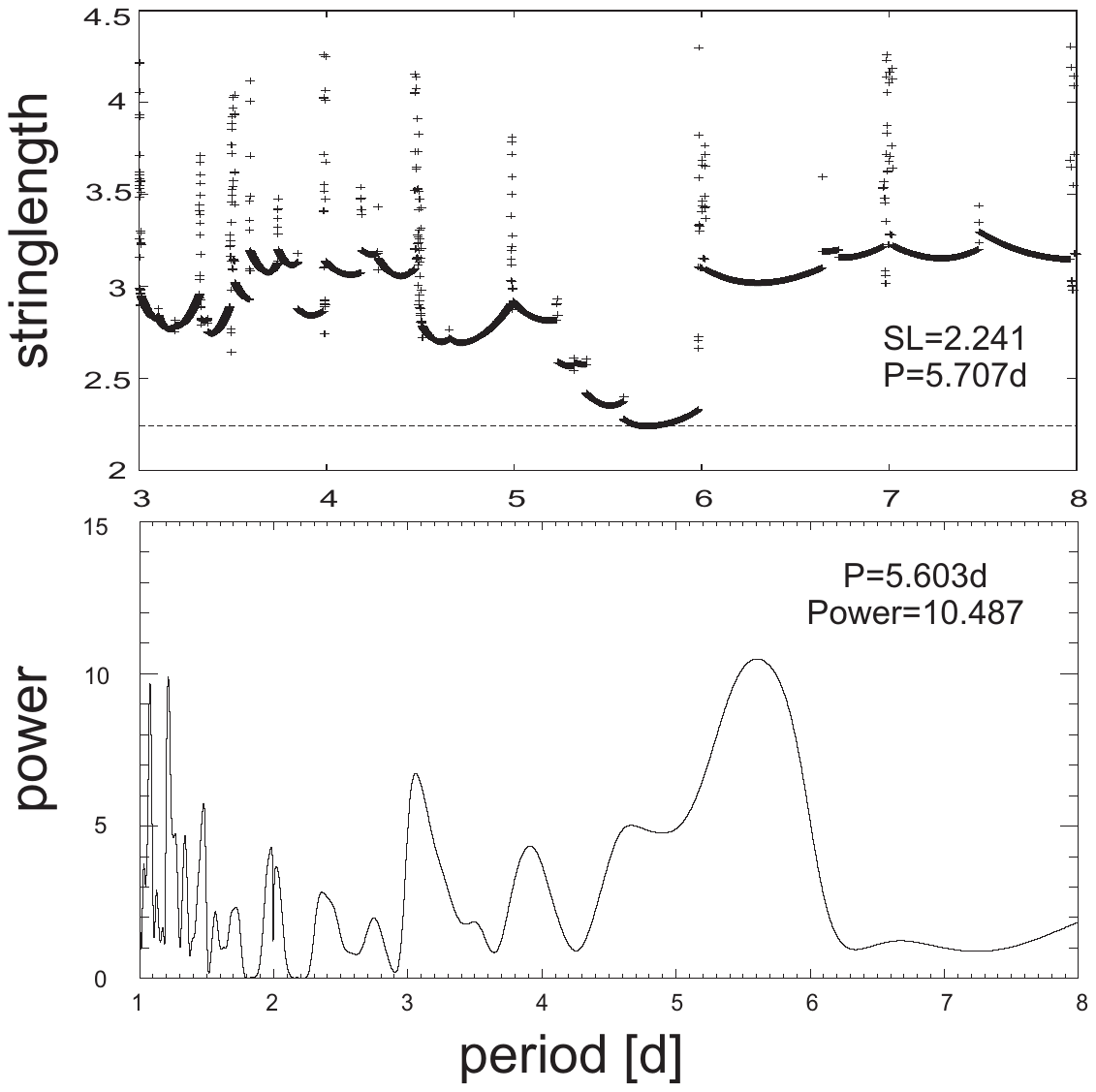}
\caption{Results of period search by the string length (\emph{top})
and Lomb-Scargle (\emph{bottom}) methods for the $V$-band data from March and
April 2007 from GSH. The minimum in the string length (2.2)  
indicates the best period being 5.71 days (\emph{top}).
The maximum power in the Lomb-Scargle test (10.5)
is for a period of 5.60 days.}
\end{figure}

\begin{figure}
\includegraphics[width=7cm,angle=0]{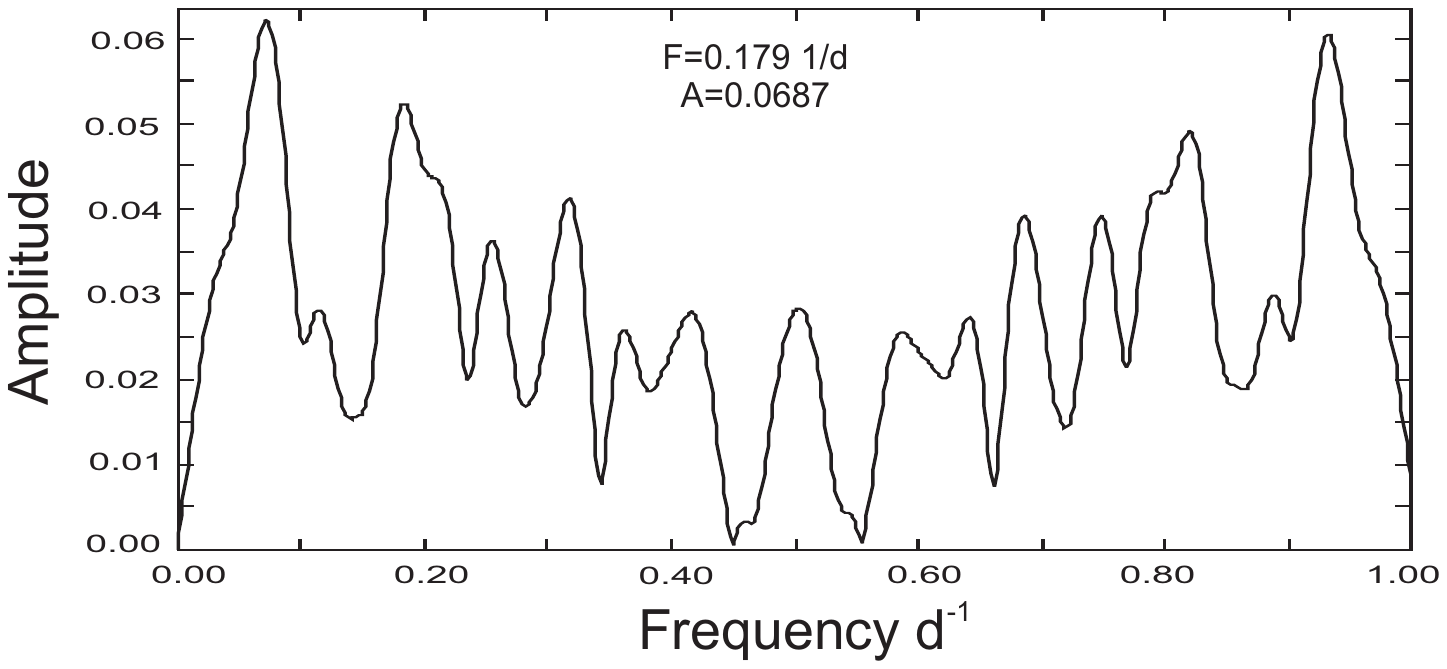}
\caption{Results of period search by Fourier transformation analysis
for the $V$-band data from March and April 2007 from GSH. 
The 1st maximum in the left corresponds to 14 days 
and is spurious due to sampling effects.
The 2nd maximum (amplitude $A$ = 0.069) from the left is for
a period of 5.59 days, which we regard as true rotational period.
The other maxima between frequencies 0.8 and 1.0 per day are alias 
signals expected (for 5.59 and 14 day periods) according to Tanner (1948).}
\end{figure}

\begin{figure}
\includegraphics[width=7cm,angle=0]{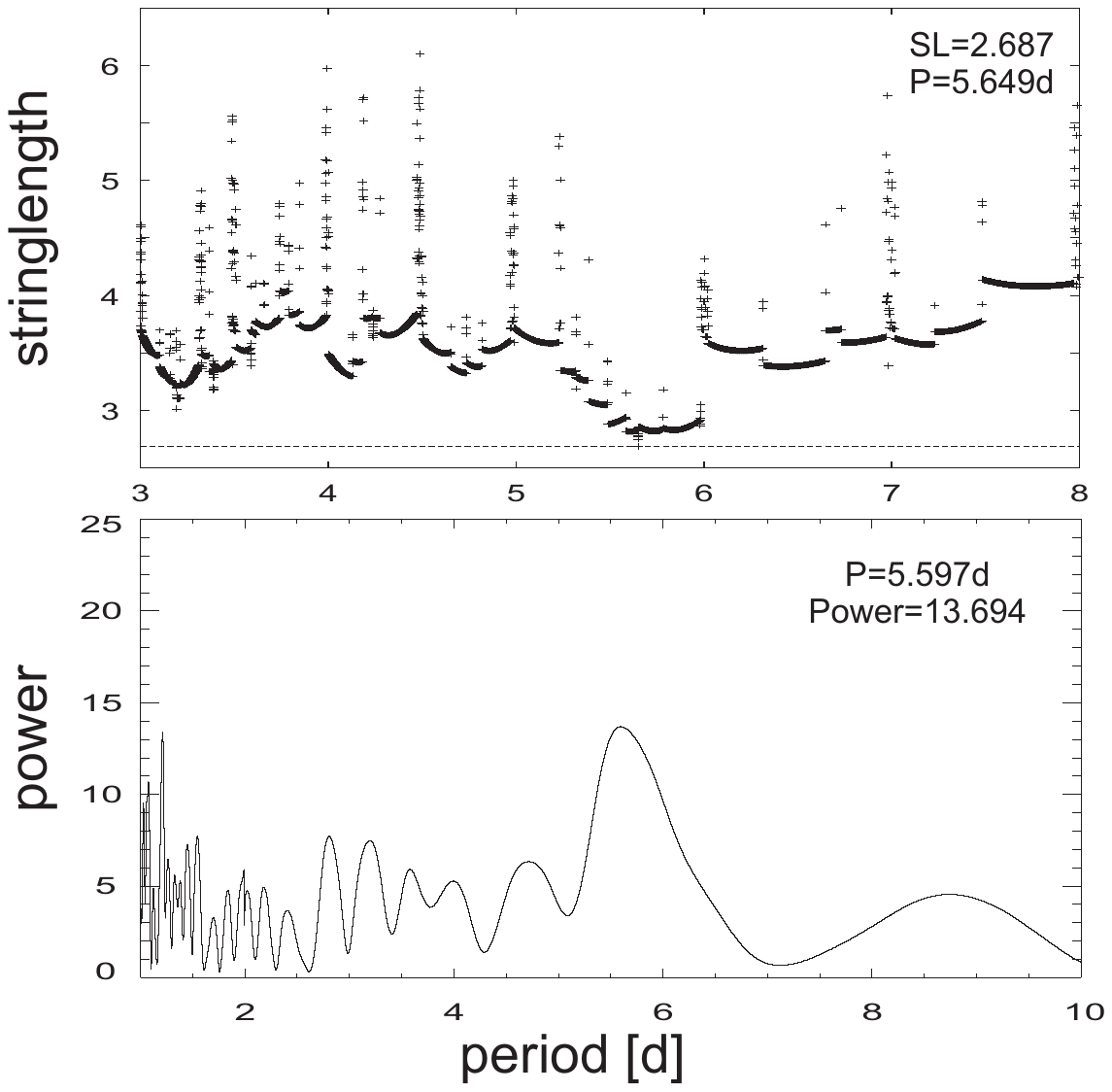}
\caption{Results of period search by the string length (\emph{top})
and Lomb-Scargle (\emph{bottom}) methods for the $R$-band data from March and 
April 2007 from GSH. The minimum in the string length (2.7) indicated the
best period: 5.65 days (\emph{top}).
The maximum power in the Lomb-Scargle test (13.7)
is for a period of 5.6 days.}
\end{figure}

%
%

\begin{figure}
\includegraphics[width=7cm,angle=0]{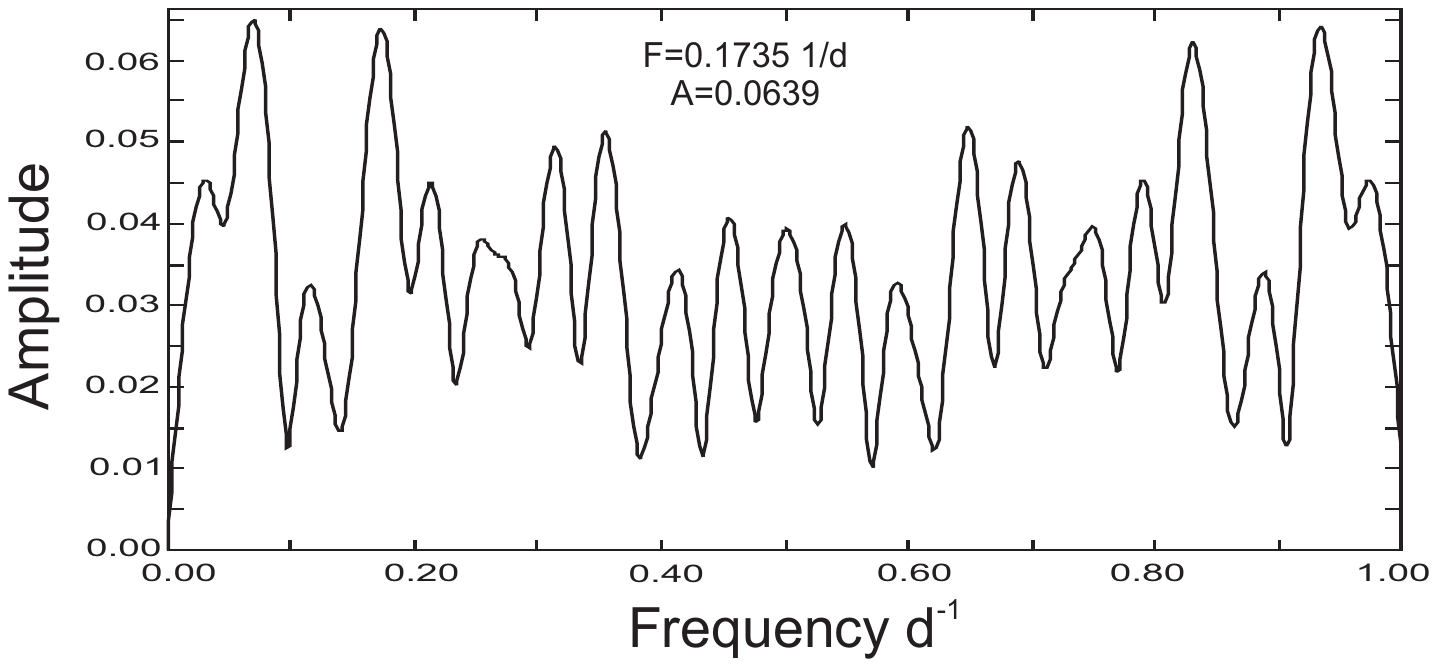}
\caption{Results of period search by Fourier transformation analysis
for the $R$-band data from March and April 2007 from GSH. The 1st maximum in the
left corresponds to 14 days and is spurious due to sampling effects.
The 2nd maximum (amplitude $A$ = 0.0639) from the left is for 
a period of 5.615 days, which we regard as true rotational period. 
The other lower maxima between frequency 0.8 and 1.0 per day are alias 
signals expected (for 5.615 and 14 day periods) according to Tanner (1948).}
\end{figure}

\begin{figure}
\includegraphics[width=5.5cm,angle=270]{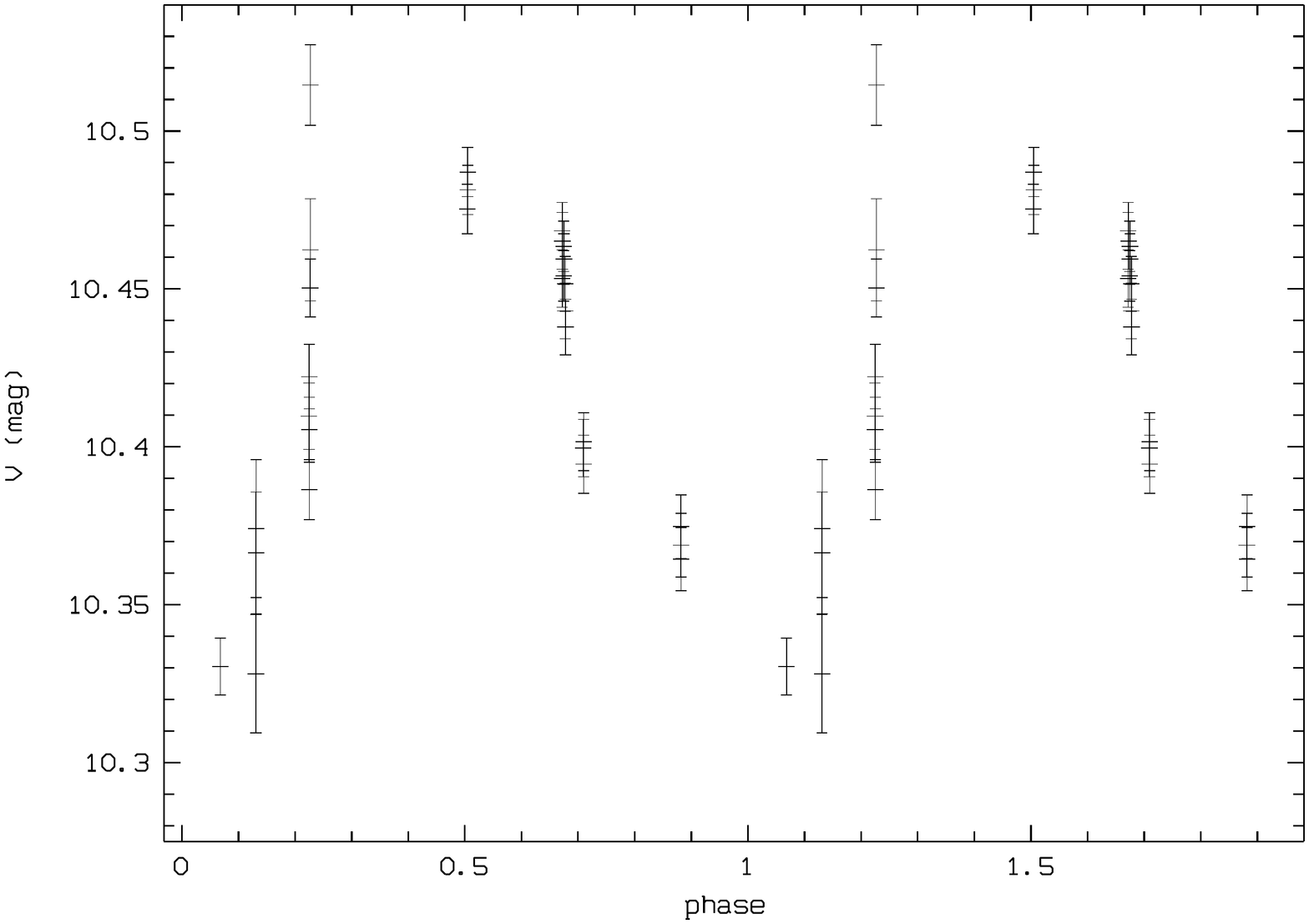}
\caption{Phased-folded plot for $V$-band photometry for Par 1724
obtained in March and April 2007 with GSH/CTK -- folded with a 5.7 day period
and showing a $\sim\! 0.2$ mag peak-to-peak amplitude.}
\end{figure}

\begin{figure}[htb]
\centering
\includegraphics[width=5.5cm,angle=270]{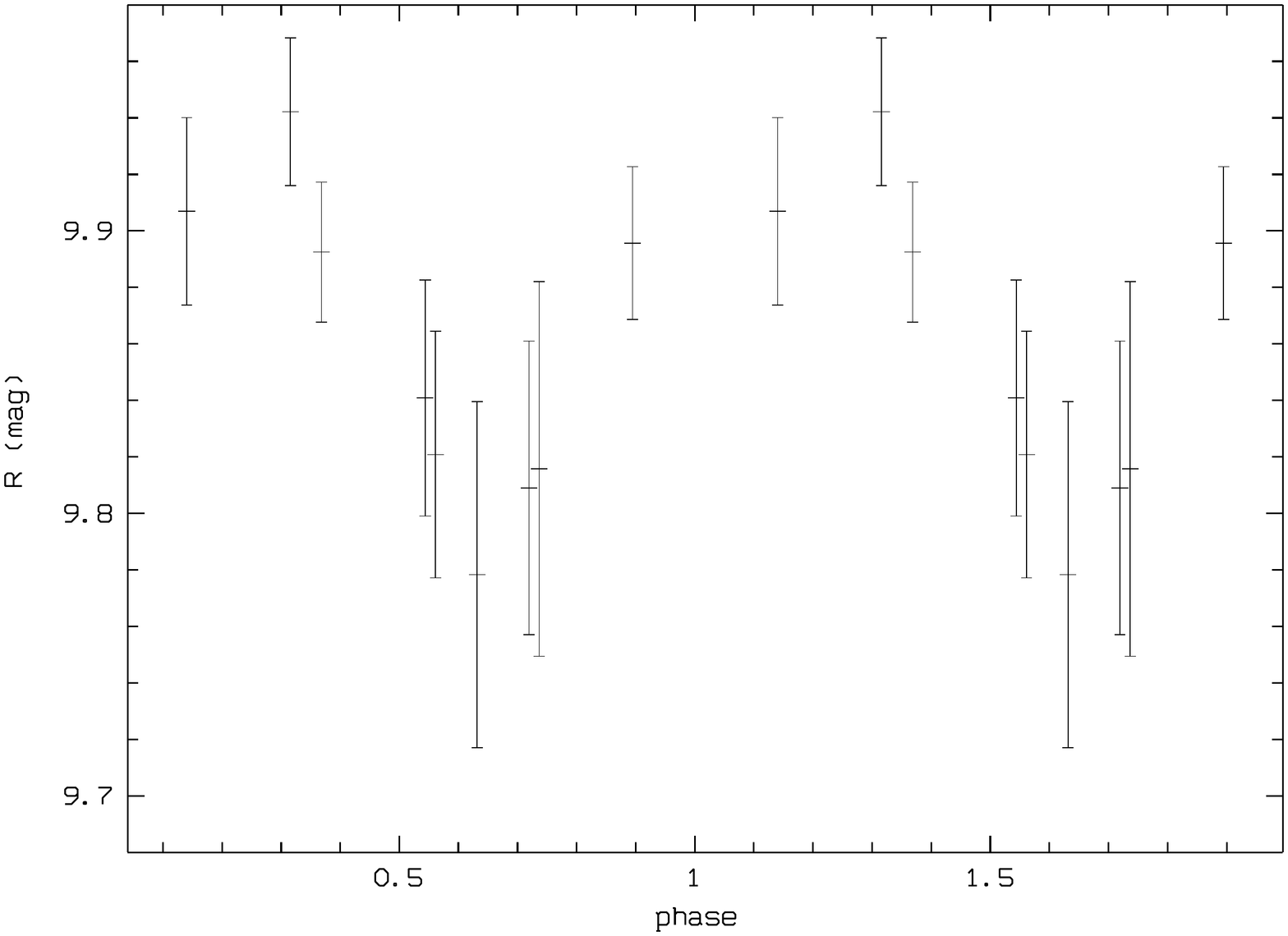}
\caption{Phased-folded plot for $R$-band photometry for Par 1724
obtained in March and April 2007 with GSH/CTK -- folded with a 5.7 day period
and showing a $\sim\! 0.15$ mag peak-to-peak amplitude.}
\end{figure}

\section{Long-term variability: a cycle?}

Using the data from the 1960ies until 1998 in Neuh\"auser et al. (1998),
those in Cutispoto (1998) and Cutispoto et al. (2001, 2003), 
and the new data listed in this work (Tables 3 and 4), we can also study 
possible long-term photometric variability. From the data of the night
we did absolute photometry, we can also derive $BV\!RI$ data points for
Par 1724 for those nearby nights (within a few rotation periods), 
when we did only relative photometry
(Table 4); those data points are included in Figs. 13 and 14 ($BV\!RI$).

In addition, we use the
$V$-band data available online\footnote{www.astrouw.edu.pl/asas}
from the All Sky Automated Survey (ASAS), where the
whole observable sky (down to 14 mag) is monitored with one 
observation per night using telescopes on Las Campanas, 
Chile, and Haleakala, Hawaii.
Almost 700 fully reduced $V$-band data points for Par 1724 are 
available (by mid Jan 2009) for five different aperture sizes;
we use the average of those five values per night (but see the 
same trend with any of the five aperture sizes). 

\begin{figure*}[htb]
\centering
\vskip-2mm
\includegraphics*[bb= 3.0cm 2cm 20cm 27cm, width=9.5cm,angle=270]
      {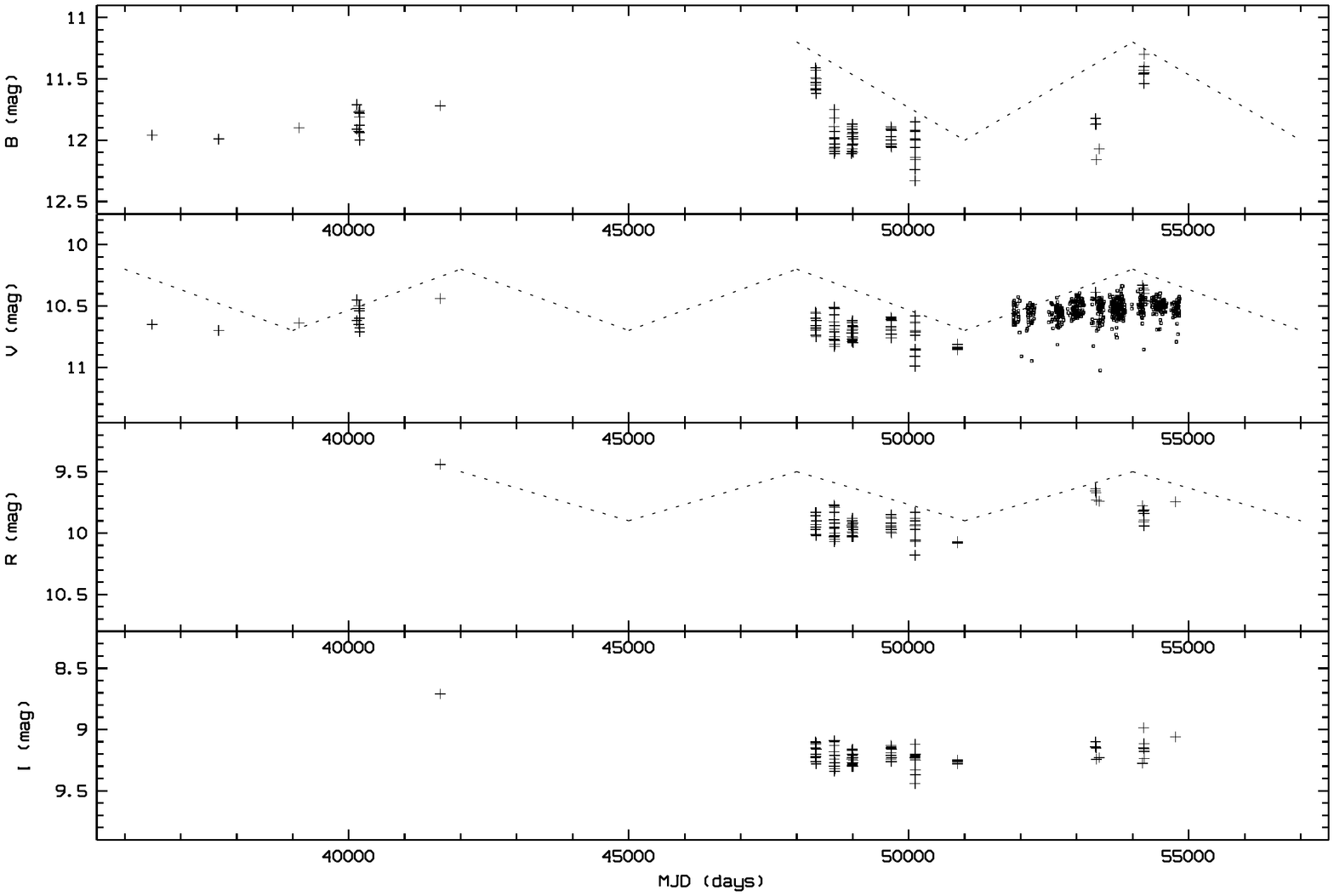}
\vskip-3mm
\caption{Absolute photometry for Par 1724 from the 1960ies until January 2009
for $B$ (\emph{top}), $V$ (\emph{second from top}), $R$ (\emph{second from bottom}), and $I$ 
(\emph{bottom}) according to data in Neuh\"auser et al. (1998), Cutispoto (1998),
Cutispoto et al. (2001, 2003), and this work (Table 4).
We plot absolute magnitudes versus Modified Julian Date
(MJD = JD -- 2400000.5).
In the plot for $V$-band data, we also show data from the 
ASAS project as small dots, see text. A possible 17.5 year cycle is
shown as upper envelope with dashed lines.
We see an increase in amplitude and a decrease in the faintest magnitudes
in $B$, $V$, $R$, and $I$ at around MJD = 50\,000.}
\end{figure*}

\begin{figure*}[htb]
\centering
\includegraphics*[bb= 4.0cm 3cm 19cm 27cm,width=8.5cm,angle=270]{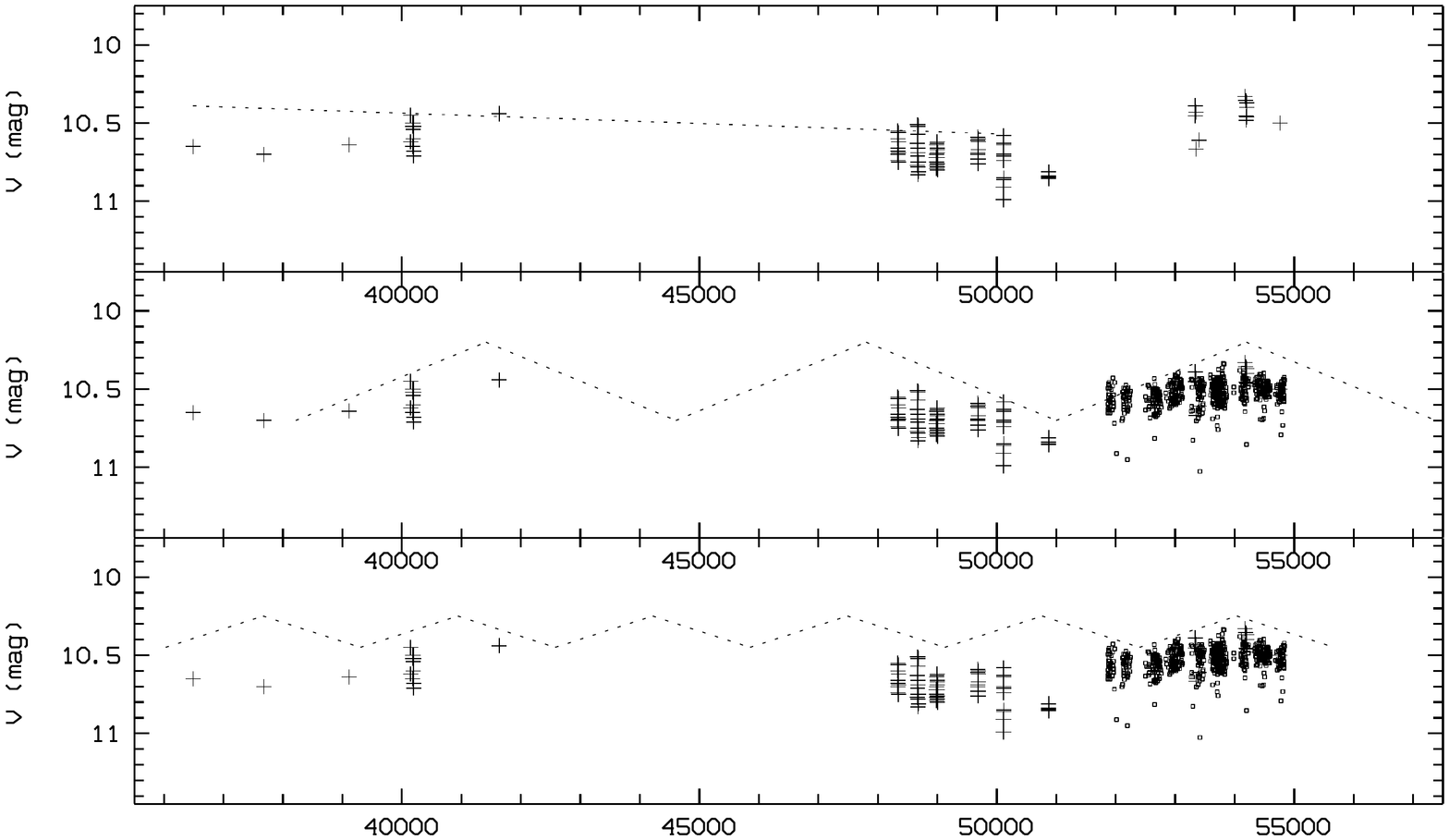}
\caption{Absolute photometry for Par 1724 from the 1960ies until January 2009
for $V$ band only,
according to data in Neuh\"auser et al. (1998), Cutispoto (1998),
Cutispoto et al. (2001, 2003), this work (Table 4),
and from the ASAS project (small dots).
We plot absolute magnitude in mag versus Modified Julian Date
(MJD = JD -- 2400000.5).
In the \emph{upper panel}, we show the small long-term trend 
suggested in Neuh\"auser et al. (1998) towards fainter magnitudes,
which we cannot confirm with the new data, Par 1724 got brighter again.
Instead of such a simple long-term trend, there may be a cyclic behaviour:
Period analysis gave 17.5 and 9 years as best periods (Fig. 15),
for which we show the suggested upper envelopes
to the $V$-band data in the \emph{middle} (17.5 year cycle) and \emph{bottom 
panel} (9 year cycle); phase-folded light curves are shown
in Fig. 16.
The data seem consistent with any of those two possible cycle length.
Also the data in the other bands ($BRI$) shown in 
Fig. 13 are consistent with such a long-term periodicity.
The possible cycle length (17.5 or 9 years) 
is similar to the 11 year solar cycle, but has to be confirmed.
The amplitude of the cycle is larger in $B$ than in $V$,
and smallest in $R$ and $I$ (Fig. 13).
In the top panel, the ASAS data are not shown just for clarity to
see the GSH, Dutch, and Wendelstein telescope data.}
\end{figure*}

\begin{figure}
\includegraphics[width=8cm,angle=0]{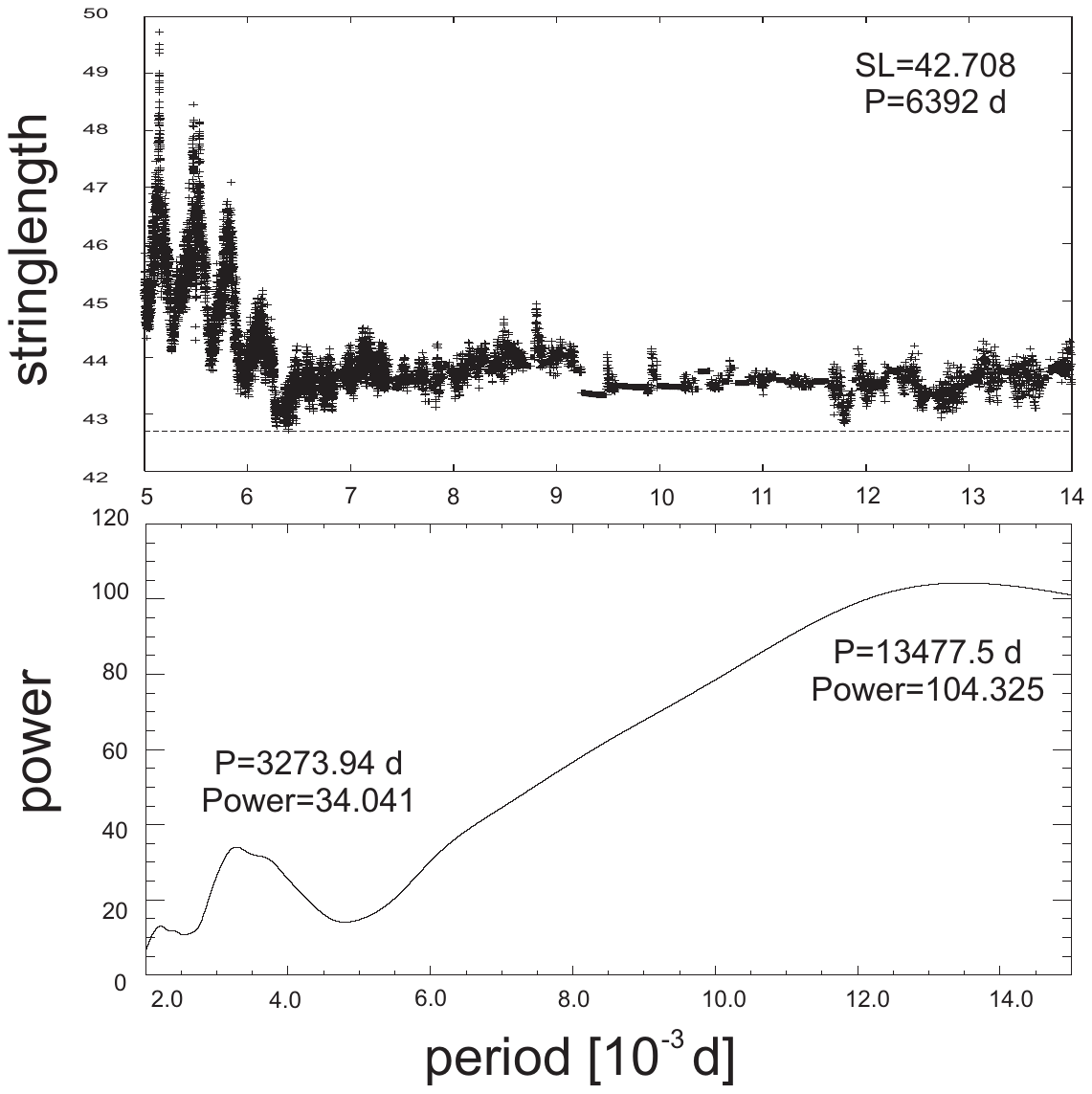}
\caption{We show here the periodgrams for the long-term variability
of Par 1724 using the ASAS $V$-band data for string length method (\emph{top})
and Lomb-Scargle (\emph{bottom}) with the possible periods (or cycle length)
of 6392 days (17.5 yr) in the top panel as well as 3274 days (9 yr) 
and 13478 days (37 yr) in the bottom panel. The most likely cycle
length are 17.5 or 9 years. Phase-folded light curves
can be found in Fig. 16. \vspace{1cm}}
\end{figure}

We used the $V$-band data of $\sim\! 40$ years to search for an
additional long-term period or cycle. The three periodogram
analysis tools used (see Sect. 4 for details) gave five different
possible long-term periods (see Fig. 15): 
\begin{itemize}
\item
10\,193.4 d = 27.9 yr with relatively high false alarm probability 
of 0.058 (Fourier); 
\item
6392 d = 17.5 yr with string length 42.7 (string length), and
\item
13\,477.5 d = 36.9 yr or 3273.9 d = 8.96 yr or 2210.93 d = 6.05 yr
with powers 104.33, 34.04, and 12.98, respectively
(Lomb-Scargle test). 
\end{itemize}
Visual inspection of the phase-folded
light curves of those five possible periods confirm that either
the 9 or 17.5 year period are possible (Fig. 16). One of them can be
an alias of the other. 
These two periods are also consistent
with the light curves plotted in Fig. 14 (magnitude versus 
observing data instead of phase).

The magnitudes in the other bands ($BRI$) as plotted in 
Fig. 13 would also be consistent with a long-term 
periodic or cyclic behaviour, a 9 or 17.5 year cycle.

Such a long-term period would be similar to the 11 year solar cycle.
Long-term brightness changes could be due to spots
getting larger or smaller in size and/or getting hotter or 
cooler in temperature and/or increasing or decreasing in number (total area).
However, we cannot draw a firm conclusion in this regard,
mainly because of the time gaps.

The $V$ band peak-to-peak amplitude was $\Delta V$\,= 0.2 to 0.3 mag 
from 1968 to 1995 (data in Neuh\"auser et al. 1998), then suddenly
increased to 0.4 mag in the season 1995/1996, as already
noticed by Cutispoto et al. (2003), then was again 0.25 mag 
from 2001 to 2006, to fall down to 0.2 in the last season 2008/2009
(Fig. 17).

In Fig. 17, we show the ASAS data of the different seasons
always folded with the 5.7 day period. We can now investigate
whether the amplitude of the variability in $V$ due to spots
changes periodically with the cycle length.
For the first five ASAS seasons (2001 to 2007), the maximum was at the same
phase (near 1) and the amplitude was also nearly constant
around 0.25 mag (peak-to-peak in $V$). Both values, however,
were different in the very last (current, incomplete)
season 2008/2009, when the amplitude is smaller (${\sim\! 0.2}$ mag).
This would indicate a decrease in spot size and/or
temperature difference, and a recent change in spot location
(i.e. a change in phase).
This should be confirmed by a new Doppler image.

We can consider whether the changes in $V$-band variability
amplitude is consistent with any of the long-term cycles suggested above.
In the solar maximum, there are more dark spots, so that the
variability amplitude would be larger.
For Par 1724, the 9 year cycle (plotted in Fig. 14) would predict
extremes in 1993 and 2001/02 (faintest magnitudes)
and in 1997/98 and 2006 (brightest magnitudes);
this is not consistent with the observations (but data gap around 1998).
The 17.5 year cycle (also plotted in Fig. 14), however, would
predict the extreme with faintest magnitudes and largest
amplitude for 1996--1999 and the extreme with brightest magnitudes 
and smallest amplitude for very recently (2005--2008).
Indeed, we saw a sudden increase in amplitude from 1995 to 1996,
where the star became as faint as ${V=11}$ mag (MJD = 50\,000), i.e. an increase in
spottedness; also in our 1998 observations at the Dutch telescope,
the star was still relatively faint (but only one night), see Table 3.
Also, we just saw a decrease in amplitude in the
last season 2008. This could be seen as supporting evidence for
the 17.5 year cycle, but has to be confirmed with further observations
and Doppler images.
The data for the previous Doppler image (Neuh\"auser et al. 1998)
were taken in December 1995 and January 1996, i.e. during the time,
when we would expect a maximum in spottedness, according to the
arguments above for a 17.5 year cycle; that Doppler image has
shown a large polar spot (or group of spots) covering $12\,\%$ of
the visible stellar disk with a temperature difference of ${\sim\! 800}$~K
compared to the surrounding photosphere (Neuh\"auser et al. 1998).
\begin{figure*}
\centering
\includegraphics*[bb= 2.5cm 2.5cm 19cm 27cm, width=9.0cm,angle=270]{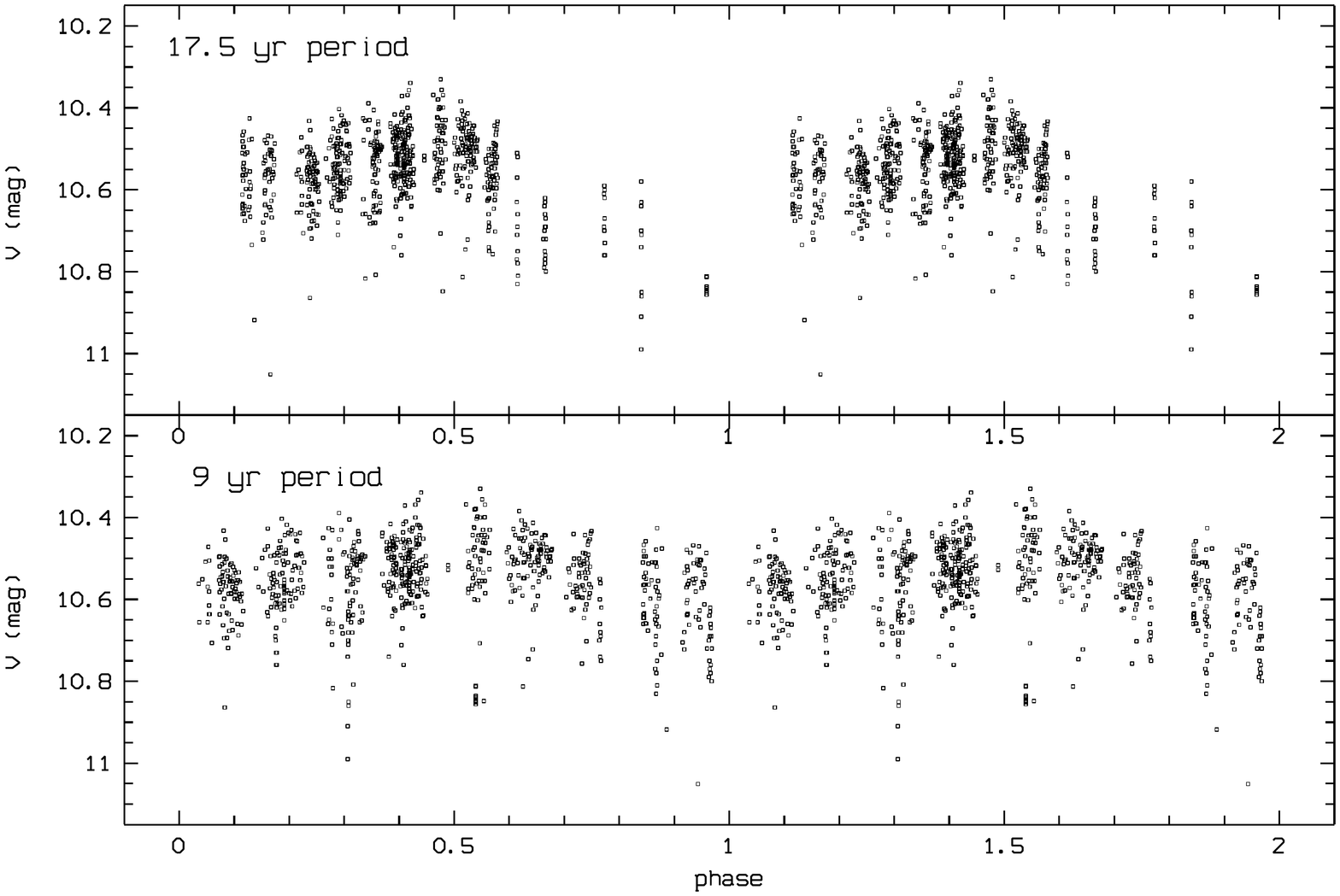}
\caption{Absolute photometry for Par 1724 from the 1960ies until January 2009
for the $V$ band only, for the same data as in Fig. 13.
We plot absolute magnitude in mag versus phase for
the two best long-term periods (or cycles), namely
for ${\sim\!17.5}$ years (6392 days) in the \emph{top panel}
and for $\sim\! 9$ years (3274 days) in the \emph{bottom panel}.
A long-term cycle would be seen not only in sinusoidal variations,
but the star should also show a smaller amplitude when being brightest
and a larger amplitude (going towards fainter magnitudes) when being
faintest; this is the case for both cycle length shown here:
For 17.5 yr (\emph{top}) the amplitude is smallest around phase 0.5 (brightest)
and largest around phase 0.85 (faintest), but with sparse data here;
for the 9 yr cycle (\emph{bottom}), this trend is even more pronounced,
small amplitude at phase 0.5 (brightest) and large amplitude 
around phase 0.8--1.0 (faintest).
The $B$-band data are also consistent with the both
17.5 yr and 9 yr cycles, while the $R$- and $I$-band data
are not inconsistent with such a cycle (but too many data gaps).
Given the results shown in Figs. 13--15, however,
it is difficult to distinguish between the two best 
possible cycles, so that further observations are needed.}
\end{figure*}
\begin{figure*}
\centering
\includegraphics[width=17cm,angle=0]{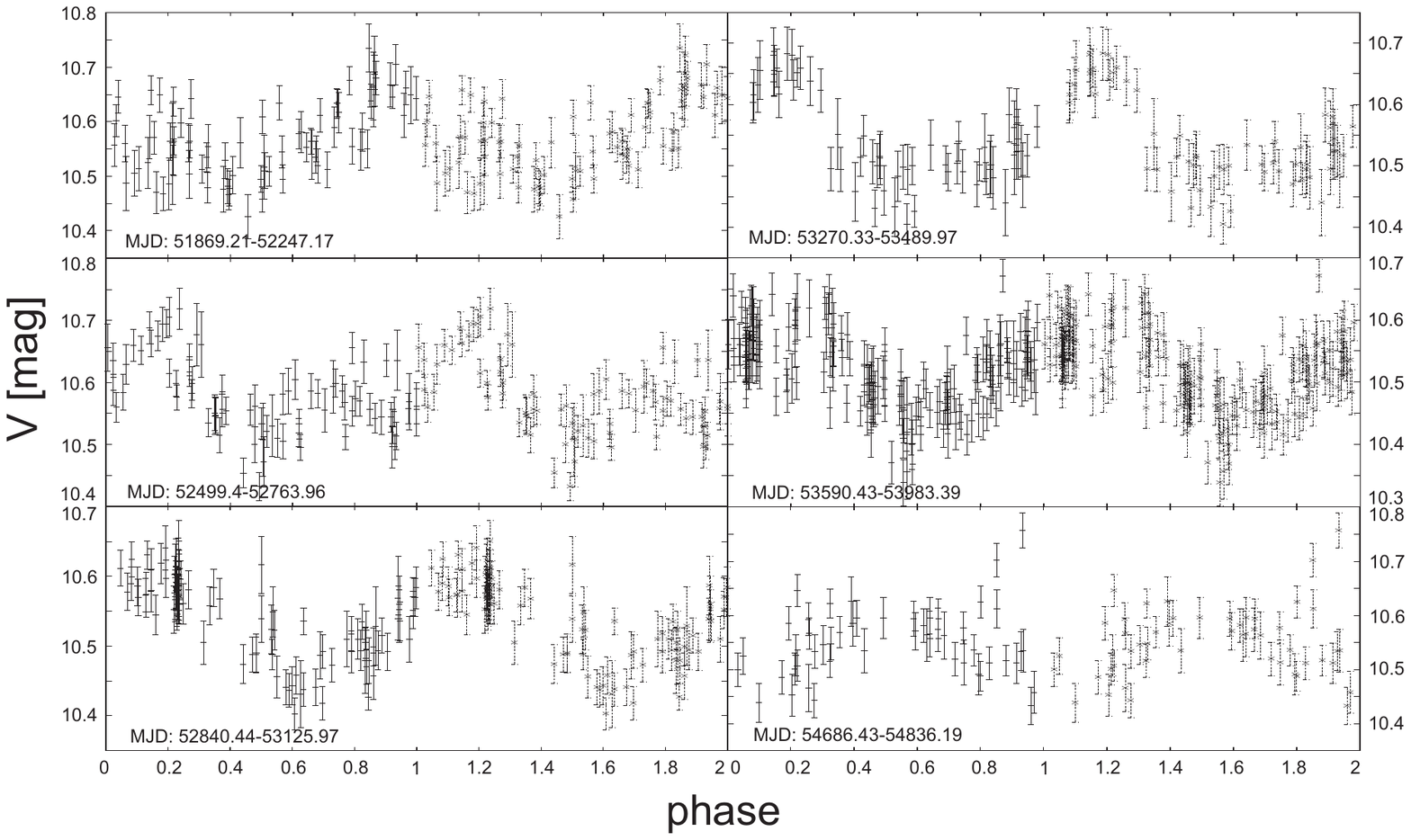}
\caption{Phase-folded light curves with ASAS $V$-band data
for the different seasons with ASAS observations
(given with MJD range), always
plotted in phase for the 5.7 day rotation period.
The maximum is always near phase 1, but different 
in the last season (lower right, most recent season 2008/2009), 
which would indicate a recent change in spot location.
The peak-to-peak amplitude is always ${\sim\! 0.25}$ mag,
but only about 0.2 mag in the last season (2008/2009),
which would indicate a recent change in spot size
and/or temperature difference.}
\end{figure*}

We will continue to monitor the star more frequently from GSH in the 
next few years and decades -- in particular also in the $B$ and $I$ bands
(not covered by ASAS) -- to confirm or reject the hypothesis of 
a long-term cycle and to constrain the cycle length.

\section{Summary}

Rotational periods and long-term cycles are important to
understand the angular momentum evolution and the dynamo
in young and active stars.

With $\sim\! 800$ new photometric observations of the young naked
weak-line run-away T Tauri star Par 1724 near the Orion Trapezium
cluster, we could confirm again the $\sim\! 5.7$ rotation period
(in the $V$ and $R$ bands), now observed in all data since the 1960ies.

In addition, with the new data from 1998 to 2009, 
we cannot confirm a steady decrease in brightness,
as was suggested in Neuh\"auser et al. (1998) with data until 1997.

However, we find for the first time indications for a 9 or 17.5 year cyclic
behaviour similar to the $2 \times 11$ year cycle in the Sun, found for Par 1724
in the $V$-band data, maybe also in $BRI$. This observation needs to be confirmed
with additional data over the next few decades, if possible also with new Doppler
imaging observations almost once every year to check for possible changes in
the spot(s) size(s), temperature distribution(s), and/or location(s).
Such a 9 or 17.5 year cycle would be similar to the cycle length found
in other relatively young active stars (Baliunas et al. 1995; Alekseev 2005),
but our star Par 1724 may be the youngest (200\,000 years) ever found 
to show a long-term brightness variability cycle.

\acknowledgement{We would like to acknowledge strong support 
from the mechanical and electronics workshop 
of the faculty for physics and astronomy at the University Jena in
putting our telescope in Gro\ss schwabhausen back to work.
We would like to thank Sabine K\"onig for participation
in some of the GSH observations.
RN would like to thank DLN for good suggestions and ideas.
We would like to thank the stuff 
at ESO La Silla for help during the observations at the 90\,cm Dutch telescope. 
For the Wendelstein observations, we would like to thank Ulrich Hopp, Heinz Barwig, 
Otto B\"arnbantner, Christoph Ries, and Wolfgang Mitch for access to the 
telescope and observational support; some of the data were taken in service mode.
RN acknowledges general support from the German National Science Foundation
(Deutsche Forschungsgemeinschaft, DFG) in grants NE 515/13-1, 13-2, and 23-1,
AK acknowledges support from the DFG in grant KR 2164/8-1,
SR and MV acknowledge support from the EU in the FP6 MC ToK project MTKD-CT-2006-042514,
TOBS acknowledges support from the Evangelisches Studienwerk e.V., Villigst,
TR acknowledges support from DFG in grant NE 515/23-1,
TE and MH acknowledge partial support from DFG in the SFB TR-7 Gravitation Wave
Astronomy, M. Moualla acknowledges support from the Syrian government.
NY acknowledges support from the UK Socrates Erasmus Council 
and the Student Loans Company. We used the Simbad and VizieR data bases.}

\end{document}